\documentclass[a4paper,11pt]{article}
\usepackage[utf8]{inputenc}
\pdfoutput=1 % if you are submitting a pdflatex (i.e. if you have
\usepackage{jheppub}
\usepackage{amsmath,amssymb,bbm,amsfonts,dsfont}
\usepackage{braket,ulem,booktabs}
\usepackage{graphicx}
\usepackage{hyperref}
\usepackage{enumitem}
\usepackage{stackrel}
\usepackage{subcaption}
\usepackage{array}
\usepackage{booktabs,tabularx}
\usepackage{tikz,tikz-cd}
\usepackage{mathdots}

\def\ra{\rightarrow}

\def\be{\begin{equation}}
\def\ee{\end{equation}}
\def\ba{\begin{eqnarray}}
\def\ea{\end{eqnarray}}
\newcommand{\smt}[1]{\smash{\widetilde{#1}}}

\def\nb{\nonumber}
\def\p{\partial}
\def\a{\alpha}
\def\b{\beta}

\def\f{\varphi}

\def\g{\gamma}
\def\G{\Gamma}
\def\d{\delta}
\def\D{\Delta}
\def\l{\lambda}
\def\L{\Lambda}
\def\m{\mu}

\def\s{\sigma}
\def\S{\Sigma}

\def\th{\theta}

\def\q{\quad}

\def\we{\wedge}
\def\dd{\mathrm{d}}

\def\mc{\mathcal}

\allowdisplaybreaks

\newcommand{\pr}[1]{\left(#1\right)}
\newcommand{\pq}[1]{\left[#1\right]}
\newcommand{\pg}[1]{\left\{#1\right\}}

\title{Stringy T-duality on the lattice and\\  the twisted Villain model
}

\author[a]{Coraline Bacq,}
\author[a]{Alessio Caddeo,}
\author[b]{Saskia Demulder,}
\author[a
]{Johanna Erdmenger}

\affiliation[a]{Institute for Theoretical Physics and Astrophysics and W\"urzburg-Dresden
Cluster of Excellence ctd.qmat, Julius-Maximilians-Universit\"at W\"urzburg,\\
Am Hubland, DE-97074 W\"urzburg, Germany}
\affiliation[b]{Department of Quantitative Methods, CUNEF Universidad,\\ Calle Almansa 101, 28040 Madrid, Spain}

 \emailAdd{coraline.bacq@uni-wuerzburg.de, alessio.caddeo@uni-wuerzburg.de, saskia.demulder@cunef.edu }

\abstract{
We address the question of whether dualities formulated in continuum field theory can be realised exactly at finite lattice spacing, rather than only emerging in the infrared. In this context, we construct a lattice framework for a genuinely stringy form of T-duality. We extend the exact lattice T-duality of the compact boson to curved backgrounds with non-trivial circle fibrations, where the duality is no longer exhausted by the familiar exchange of momentum and winding, but also involves global topological data. To this end, we define the twisted Villain model, which couples the lattice fibre field to cochains encoding the bundle connection and the fibre-horizontal component of the $B$-field. We realise this structure in lattice models for several fibred backgrounds and recover the characteristic bundle-flux exchange of T-duality. Using a half-gauging procedure, we derive the associated lattice defect action and show that it defines a topological defect. This establishes that the distinctive topological features of T-duality on curved manifolds can be captured exactly in a lattice model, implying that this duality is not tied to a particular continuum representation is present in lattice-regularised models. }

\begin{document}

\maketitle

\newpage
%%%%%%%%%%%%%%%%%%%%%%%%%%%%%%%%%%%%%%%%%%%%%%
%%%%%%%%%%%%%%%%%%%%%%%%%%%%%%%%%%%%%%%%%%%%%%
\section{Introduction}

Many dualities are most naturally formulated in continuum field theory, where they reveal unexpected relations between apparently different descriptions of the same physics. A natural question is then how much of this structure survives in a finite-spacing lattice setting. In particular, when a continuum theory arises as the long-distance limit of a lattice model, one may ask whether its characteristic dualities are merely emergent in the infrared, or whether they can already be realised exactly on the lattice. 
 The latter possibility is especially interesting, not only because the lattice provides a finite-spacing, discrete definition of the theory, but because an exact lattice realisation shows that the duality is not tied to a particular continuum presentation. Rather, it isolates the part of the duality that is sufficiently robust to survive discretisation.

A canonical example of such a duality is the T-duality of the compact boson, which exchanges momentum and winding and relates theories at radii $R$ and $\alpha'/R$. Standard lattice systems such as the XY rotor model \cite{Jose:1977gm} and related spin chains \cite{Lieb:1961fr}, including the XXZ spin chain \cite{Baxter:1982zz}, reproduce this theory in the infrared. In many important cases, however, this route is not quite sufficient. The continuum theory may exhibit symmetries that are only emergent at long distances, or dualities that become exact only in the infrared.  In the compact-boson case, the ordinary Villain \cite{villain1975theory} or rotor formulations flow to the free boson only in an appropriate regime \cite{Alcaraz:1987ix,Baake:1988zk}, and the winding symmetry is not exact in these lattice models. By contrast, the modified Villain construction \cite{Sulejmanpasic:2019ytl,Gorantla:2021svj} improves this situation precisely by imposing flatness on the integer gauge field. In this way, the continuum momentum and winding symmetries become exact already on the lattice, and the model realises exact lattice T-duality rather than merely an infrared relation. As shown in \cite{Gorantla:2021svj,Cheng:2022sgb,Argurio:2026txf,Aoki:2026pvq}, modified Villain constructions are especially well suited to making such structures precise in a lattice system. A related earlier example is \cite{Gross:1990ub} which showed that vortices obstruct the naive $R\rightarrow \alpha'/R$ duality, while a vortex-free definition restores it. More recently, exact lattice T-duality was also realised in the XX-spin chain through a non-invertible symmetry exchanging lattice momentum and winding symmetries \cite{Pace:2024oys}. One can therefore do better than simply engineering the correct continuum limit, and construct lattice theories in which the relevant symmetry and duality structures are already exact at finite lattice spacing.

The compact-boson example already suggests the broader question of whether exact lattice realisations persist once T-duality is applied in more genuinely geometric settings. Indeed, T-duality becomes richer in curved backgrounds. In the standard Buscher formulation of the duality \cite{Buscher:1987qj,Rocek:1991ps,Alvarez:1994dn}, the relevant ingredient is not an exact global winding symmetry, but a circle isometry generated locally by a Killing vector. When this Killing vector generates a non-trivially fibred $U(1)$, T-duality generically no longer acts simply by exchanging momentum and winding charges. Rather, it exchanges topological and flux data: for a circle bundle with $H$-flux, the topology of the dual bundle is determined by the original flux, while the dual flux is determined by the Chern class of the original fibre-bundle \cite{Bouwknegt:2003wp,Bouwknegt:2003vb}. The simplest and most familiar example is the T-duality of the Hopf fibre of $S^3$. Earlier continuum constructions of T-duality walls and defects appear in \cite{Sarkissian:2008dq,Kapustin:2009av,Bachas:2012bj}, while more recently non-invertible T-duality defects in curved sigma-model settings have been constructed in \cite{Demulder:2022nlz,Bharadwaj:2024gpj,Arias-Tamargo:2025xdd}. This is therefore a genuinely string-theoretic form of T-duality, rather than a phenomenon tied only to the free compact boson. 

In the present work, we study this problem in the language in which it is most naturally formulated, namely that of non-linear sigma models, but with a different question in mind from the usual continuum-limit one. We ask not merely for a lattice model whose long-distance limit reproduces a given sigma model, but whether the duality itself can  be made exact at finite lattice spacing. In this way, the lattice serves not simply as an approximation scheme for the continuum theory, but as a test of which topological and symmetry data are genuinely intrinsic to the duality and survive discretisation. Lattice systems may realise continuum symmetries and anomalies in ways that are exact, emergent, or visible only in the infrared. Hence the existence of a desired continuum limit alone does not guarantee an exact lattice duality structure; see \cite{Cheng:2022sgb}.

This raises the central question of the present paper: can the topological content of stringy T-duality be made exact directly on the lattice? Our answer is that it can. 
More concretely, we show that this topological form of T-duality admits an exact lattice realisation without requiring a complete lattice realisation of the full curved sigma-model. The key simplification is that, although fibred T-duality arises in backgrounds with non-trivial geometry and topology, the data exchanged by the duality are carried by the Abelian fibre sector: the circle-bundle curvature, and the fibre-horizontal component of the $B$-field. This allows us to formulate the problem entirely in terms of lattice cochains, avoiding the need for non-Abelian lattice variables. Within this framework, we construct lattice avatars of several non-linear sigma models and show that they enjoy an exact T-duality featuring the characteristic bundle-flux exchange of topological data. Following the half-gauging procedure \cite{Choi:2021kmx,Choi:2022zal}, we derive the action of the non-invertible T-duality defect and establish its topological nature. An important consequence is that, beyond the free compact boson, the exact realisation of T-duality no longer relies on an exact global winding symmetry, but instead on the appropriate bundle and flux data, in agreement with the continuum Buscher and topological-T-duality picture.

A natural language for organising this exact lattice realisation is that of generalised symmetries \cite{Gaiotto:2014kfa} and associated topological defects. This framework provides a common way of describing symmetries and dualities in both continuum and lattice systems. In this language, dualities can be encoded by topological defects, often of non-invertible type. T-duality defects have appeared in several related settings. In the free-boson and conformal-interface context they were identified as distinguished topological defects between dual compact-boson theories \cite{Fuchs:2007tx}. In a more geometric and sigma-model language, duality walls and T-duality defects were further developed in \cite{Sarkissian:2008dq,Kapustin:2009av,Kapustin:2010if,Bachas:2012bj,Niro:2022ctq}. More recently, these structures have been reinterpreted from the viewpoint of non-invertible symmetries \cite{Thorngren:2021yso,Choi:2021kmx,Choi:2022zal,Pace:2024oys}, and extended to broader classes of T-duality defects and related sigma-model settings \cite{Bharadwaj:2024gpj,Demulder:2022nlz,Arias-Tamargo:2025xdd,Arias-Tamargo:2025fhv}.

There are by now several discrete approaches to stringy systems, ranging from random-surface and triangulation constructions to discretised worldsheets \cite{Bliard:2022oof}, lattice discretisations of gauge-fixed worldsheet theories in fixed curved backgrounds \cite{Forini:2016gie,Forini:2016sot}, and discrete descriptions of classical string motion such as segmented strings in $\mathrm{AdS}_3$ \cite{Callebaut:2015fsa,Gubser:2016zyw,Vegh:2021jhl}. These directions are close in spirit insofar as they also seek finite or discrete formulations of string-theoretic systems, however the focus here is complementary. Those works aim either at ap\-proximating the full worldsheet dynamics or at recovering a continuum string theory from discrete microscopic data. Here however, we do not attempt a full lattice discretisation of the non-linear sigma model or of the complete target-space dynamics. Instead, building on the exact lattice T-duality of modified-Villain-type compact-boson systems \cite{Fazza:2022fss,Gorantla:2021svj}, we isolate the Abelian fibre sector and the cohomological data that control fibred T-duality and realise them exactly at finite lattice spacing. In this sense, our construction provides a lattice realisation of the topological sector of stringy T-duality.

\medskip

The paper is organised as follows. 
In section \ref{sec:reviewTduality}, we review continuum T-duality for circle fibrations and its topological formulation in terms of bundle-flux exchange and correspondence-space descent. 
In section \ref{sec:twistedVillain}, after recalling the ordinary modified Villain model, we introduce the twisted Villain model for non-trivially fibred $U(1)$ backgrounds and analyse its symmetry structure.
In section \ref{sec:explicitexamples}, we illustrate the construction on several examples, including the torus with $H$-flux, the nilfold and the fiber-component of the Hopf sphere. 
In section \ref{sec:defect}, we implement lattice stringy T-duality by half-gauging, derive the corresponding defect action, and prove its topological nature. We conclude in section \ref{sec:conclusion}.

%%%%%%%%%%%%%%%%%%%%%%%%%%%%%%%%%%%%%%%%%%%%%%
%%%%%%   T-duality and top. exchange    %%%%%%
%%%%%%%%%%%%%%%%%%%%%%%%%%%%%%%%%%%%%%%%%%%%%%
\section{Review: T-duality and topological exchange}
\label{sec:reviewTduality}

We begin with providing an overview of the main features of T-duality in non-linear sigma models \cite{Buscher:1987qj, Hull:2006qs}, focusing on the aspects that we aim to reproduce at the lattice level.

%%%%%%%%%%%%%%%%%%%%%%%%%%%%%%%%%%%%%%%%%%%%%%
%%%%%%%%%%%%%%%%%%%%%%%%%%%%%%%%%%%%%%%%%%%%%%
\subsection{Non-linear sigma models and T-duality}
\label{seC:nonlinearsigmamodels}
Let us consider a worldsheet $\Sigma$, on which we give coordinates $\sigma^{a}$, and a target space $E$, on which we give coordinates $X^{i}$.\footnote{We work in Euclidean signature.} Non-linear sigma models are theories of fields $X^{i}(\s)$ with action  
\be
\label{eq:NLSMaction}
S = \frac{1}{2} \int_{\S} g_{ij} d X^{i}  \we * d X^{j} - \frac{i}{2}   \int_{\S} B_{ij} d X^{i} \we d X^{j} \ .
\ee
We use $d$ for the exterior derivative on the worldsheet and $\dd$ for the exterior derivative on the target space $E$. Here, $g_{ij}$ is the metric of the target space and $B_{ij}$ the Kalb-Ramond field. The second term in (\ref{eq:NLSMaction}), called Wess-Zumino term, is, in fact, better defined as the integral of (the pullback of) $H=\dd B$ over a three-dimensional manifold $\Omega$ whose boundary is the worldsheet, $\p \Omega = \S$. The theory is independent of the choice of $\Omega$ provided that the flux of $H$ is quantised.

The simplest theory in the class (\ref{eq:NLSMaction}) is the compact boson model, where the target space is a circle $E=S^{1}$ with radius $R$ and the action is 
\be
S = \frac{R^{2}}{4 \pi} \int _{\S} d \Phi \we * d \Phi \ , \q \q \q \Phi \sim \Phi + 2 \pi \ .
\ee
The model features a $(U(1)_m\times U(1)_w)\rtimes \mathbb Z_2$ global symmetry, where the two $U(1)$ factors are referred to as momentum and winding symmetry. The theory displays T-duality, which exchanges momentum and winding symmetries and maps the radius $R$ to its inverse $1/R$. A simple way to show the duality is to introduce a background gauge field $C$ for the $U(1)_{m}$ momentum symmetry, and a compact field $\widetilde \Phi$ whose path-integration enforces the flatness of $C$ \cite{Buscher:1987qj, Rocek:1991ps, Alvarez:1994dn}. Integrating out $\widetilde\Phi$ returns the original model, while integrating out $C$ yields the T-dual theory.

T-duality is not restricted to the compact boson model; it can be established for a broad class of non-linear sigma models \cite{Buscher:1987qj, Rocek:1991ps, Hull:2006qs}. It is rooted in the existence of isometries of the target space $E$.
Let $k$ be a nowhere-vanishing Killing vector generating a free $U(1)$-action on the target space $E$. Then $E$ is a principal circle bundle over a base $M$,
\begin{align}
S^1 \hookrightarrow E \xrightarrow{\;\pi\;} M \ .
\end{align}
Choosing coordinates $X^{i}=(Y^\mu,\Phi)$ on $E$, where $Y^{\m}$ are coordinates on $M$ and $\Phi \in [0, 2 \pi)$ is a coordinate along the fibre direction (so that $k = \p_{\Phi}$), one can write the metric as
\begin{align}
\label{eq:decompmetric}
g = g_{{\rm bas}} + \frac{R^{2}}{2 \pi} \xi\otimes \xi \ , \q \q \q \xi \equiv \dd \Phi - A\ .
\end{align}
Here,  $g_{{\rm bas}}$ is a metric on the base $M$, $R$ is the fibre radius, and $A=A_\mu(Y) \dd Y^\mu$ is a connection one-form on the circle bundle. Forms that depend only on the base coordinates and that have non-vanishing components only along the base directions are called \textit{basic}, thus $A$ is a basic form.
Its exterior derivative yields the curvature of the bundle $F  \equiv \dd A = - \dd \xi$, thus fixing a non-trivial Chern-class when the fibration is non-trivial, 
\begin{align}
\frac{1}{2\pi}[F]=c_1(E)\in H^2(M,\mathbb Z)\ .
\end{align}

Non-linear sigma models feature,  in addition, a closed three-form $H$ that must be invariant under the isometry,  $\mathcal L_k H = 0$, where $\mc{L}_{k}$ denotes the Lie derivative along the vector field $k$.
The fibrewise contraction of $H$ is then a closed basic two-form, %and we assume that it is exact,
\be
\label{eq:iotaH_v}
\iota_k H = \frac{1}{2\pi} \dd V \equiv \frac{1}{2\pi} \widetilde F \ ,
\ee
for some basic one-form $V=V_\mu(Y)\dd Y^\mu$. One can then decompose $H$ as
\be
\label{eq:H_decomp2}
H = H_{{\rm bas}} + H_{{\rm fib}} \ , \q \q \q
H_{{\rm fib}} = \frac{1}{2 \pi} \dd V \we \dd \Phi \ ,
\ee
where $H_{{\rm bas}}$ is a basic closed form. Since $H = \dd B$, we can decompose $B$ as well, 
\be
\label{eq:B_decomp}
B = B_{{\rm bas}} + B_{{\rm fib}} \ , \q \q \q B_{{\rm fib}} = \frac{1}{2\pi} V \we \dd \Phi \ , 
\ee
where $\dd B_{{\rm bas}} = H_{{\rm bas}}$. Equivalently, in terms of $\xi$ defined in (\ref{eq:decompmetric}),
\be
\label{eq:H_decomp}
H =  H_{{\rm bas}} ' + \frac{1}{2\pi}\,\widetilde F \wedge \xi   \ , \q \q \q   H'_{{\rm bas}} = \dd B_{{\rm bas}} + \frac{1}{2 \pi} \tilde F \we A \ .
\ee
Notice that $H'_{{\rm bas}}$ is basic but not closed.

According to these decompositions, the non-linear sigma model action (\ref{eq:NLSMaction}) can be rewritten as
\begin{subequations}
\label{eq:NSLMaction2}
\begin{align}
\label{eq:action_base_fiber}
    S=S_{\rm{bas}}+\frac{R^2}{4\pi}\int_{\S} ( d \Phi - A)\wedge \star ( d \Phi - A) - \frac{i}{2\pi}\int_{\S} V \wedge  d \Phi \ , 
\end{align}
where
\begin{align}
S_{\rm bas}=\frac{1}{2}\int_\Sigma \pr{g_{{\rm bas}}}_{\mu\nu} d Y^\mu\wedge \star d Y^\nu - \frac{i}{2}\int_\Sigma \pr{B_{{\rm bas}}}_{\mu\nu}d Y^\mu\wedge  d Y^\nu\ .
\end{align}
\end{subequations}
Also for non-linear sigma models, T-duality can be derived following the steps outlined for the compact boson, namely gauging the isometry generated by $k$, introducing a worldsheet gauge field $C$, and a dual scalar $\widetilde\Phi$ that imposes the flatness of $C$ \cite{Buscher:1987qj, Rocek:1991ps, Alvarez:1994dn}. Depending on the order of path integration, one retrieves the original or the dual non-linear sigma model.

The non-linear sigma models (\ref{eq:NSLMaction2}) are characterised by global symmetries that depend on topological data. We mentioned above that in the topologically-trivial case of the compact boson model, the theory exhibits two $U(1)$ symmetries, namely the momentum and the winding symmetry. For generic non-linear sigma models, we still have two symmetries that we call  \textit{isometry symmetry} and  \textit{dual isometry symmetry}.  If the Chern numbers of $\tilde F$ and $F$ are, respectively, $\kappa$ and $\l$, the isometry symmetry and the dual isometry symmetry are $\mathbb Z_{\kappa}$ and $\mathbb Z_{\lambda}$ \cite{Arias-Tamargo:2025fhv}.\footnote{In fact, when the base manifold $M$ admits several 2-cycles, for each of them we have a different $\kappa_i$ and $\lambda_i$. In this case, $\kappa$ and $\lambda$ denote, respectively, the greatest common divisor of the $\kappa_i$ and the $\lambda_i$.}

%%%%%%%%%%%%%%%%%%%%%%%%%%%%%%%%%%%%%%%%%%%%%%
%%%%%%%%%%%%%%%%%%%%%%%%%%%%%%%%%%%%%%%%%%%%%%
\subsection{Bundle-flux exchange in fibred T-duality}
\label{sec:bundle-flux-exchange}

A distinctive feature of T-duality in the presence of a non-trivial circle fibration is that it exchanges bundle and flux data \cite{Bouwknegt:2003vb,Bouwknegt:2003wp}; see also \cite{Bunke:2005sn,Evslin:2008zm}. This is made particularly transparent in the topological formulation of fibred T-duality. When the fibrewise contraction $\iota_k H$ is only locally exact, the dual one-form $V$ should be understood globally as a connection on a dual $U(1)$ bundle. In this way, T-duality relates a principal $S^1$ bundle with $H$-flux to another circle bundle over the same base, with the topology of the dual bundle determined by the original flux, and conversely the dual flux determined by the curvature of the original bundle.

In the present paper, we use this framework as the clearest way to isolate the bundle-flux exchange underlying fibred T-duality, and to identify the part of the continuum structure that is retained in the lattice construction. Consider therefore a background in which the target space $E$ is a principal $S^1$ bundle over a base manifold $M$,
\begin{align}
S^1 \hookrightarrow E \xrightarrow{\pi} M\ ,
\end{align}
with first Chern class
\begin{align}\label{eq:c1}
c_1(E)=\frac{[F]}{2 \pi}\in H^2(M,\mathbb Z)
\end{align}
and supporting a closed three-form flux
\begin{align}\label{eq:int_H_class}
\frac{H}{2 \pi}\in H^3(E,\mathbb Z)\ .
\end{align}
T-duality relates this background to another circle bundle $S^1 \hookrightarrow \widetilde E \xrightarrow{\hat\pi} M$, equipped with a dual flux $\widetilde H /2 \pi\in H^3(\widetilde E,\mathbb Z)$, such that
\begin{align}\label{eq:topol_exchange}
c_1(\widetilde E)=\pi_*\left( H\right)\ , \qquad c_1(E)=\hat\pi_*\left( \widetilde H\right)\ .
\end{align}
Equivalently, if $\widetilde F$ denotes the curvature of the dual circle bundle, then
\begin{align}
\label{eq:relationHandtildeF}
\frac{1}{2\pi}[\widetilde F]=\pi_*\!\left( [H]\right)\ , \qquad \frac{1}{2\pi}[F]=\hat\pi_*\!\left( [\widetilde H]\right)\ .
\end{align}

Since $H$ and $\widetilde H$ live on different total spaces, they can be compared only after pullback to the correspondence space
\begin{align}
E\times_M \widetilde E= \{(x,\tilde x)\in E\times \widetilde E \;|\; \pi(x)=\hat\pi(\tilde x)\}\ .
\end{align}
Denoting the projections by $p:E\times_M \widetilde E\to E$ and $\hat p:E\times_M \widetilde E\to \widetilde E$, one may choose global connection one-forms $\mathcal A$ on $E$ and $\widetilde{\mathcal A}$ on $\widetilde E$, normalised so that $\pi_*\mathcal A=1$ and $\hat\pi_*\widetilde{\mathcal A}=1$. We use the notation $\mathcal A$ and $\widetilde{\mathcal A}$ here to emphasise that these are global connection one-forms on the two circle bundles, before choosing a local trivialisation such as in \eqref{eq:decompmetric}. The pulled-back fluxes then obey the relation
\begin{align}\label{eq:Poincare_descent}
\mathrm d_{\mathcal C}\mathcal P= -p^\ast H+\hat p^\ast\widetilde H\ , \qquad \mathcal P\equiv p^\ast \mathcal A\wedge \hat p^\ast\widetilde{\mathcal A}\ ,
\end{align}
where $\mathrm d_{\mathcal C}$ is the exterior derivative on the correspondence space. Thus the pullbacks of the two $H$-fluxes are cohomologous on $E\times_M\widetilde E$. For us, the important point is not the full topological-T-duality formalism built around $\mathcal P$, but the fact that the bundle-flux exchange admits such a local correspondence-space descent description.

In the local decomposition \eqref{eq:H_decomp}, the two-form $\widetilde F=\dd V$ is the basic curvature extracted from the fibrewise contraction of $H$; here $V$ is a local basic potential, not the global connection one-form $\widetilde{\mathcal A}$ introduced above. In particular, the pair of basic curvatures already captures the topological exchange relevant for dualisation along the circle fibre, namely
\begin{equation} \label{eq:topologicalexchange}
F=\mathrm d A \qquad \longleftrightarrow \qquad \widetilde F=\mathrm d V \ .
\end{equation}

For our purposes, it is primarily this bundle-flux exchange, together with the associated symmetry pattern, that will be retained in the lattice model. The r\^ole of the lattice construction we define in the next section is not to reproduce the full topological-T-duality formalism, but to capture the fibre-sensitive topological data responsible for the bundle-flux exchange of fibred T-duality.

%%%%%%%%%%%%%%%%%%%%%%%%%%%%%%%%%%%%%%%%%%%%%%
%%%%%%%%%%%%        Villain       %%%%%%%%%%%%
%%%%%%%%%%%%%%%%%%%%%%%%%%%%%%%%%%%%%%%%%%%%%%
\section{Lattice $U(1)$-fibrations and the twisted Villain model}
\label{sec:twistedVillain}

In this section, we construct the twisted Villain model, which is the elementary lattice action needed to capture fibred T-duality at finite lattice spacing. Our starting point is the modified Villain model \cite{Sulejmanpasic:2019ytl,Gorantla:2021svj}, whose exact lattice T-duality provides the basic microscopic structure that will be twisted by bundle and flux data.

We first recall in section \ref{eq:reviewmodifiedVillain} the aspects of the modified Villain model that are essential for our construction. We then introduce in section \ref{sec:twistedVillainmodel} the twisted Villain model, which incorporates the lattice data encoding non-trivial fibration and fibre-horizontal flux. Finally, in section \ref{sec:symm_tV}, we analyse its global symmetries and compare them both with those of the modified Villain model and with the symmetry structure of continuum fibred non-linear sigma models. In this way, we make precise how the familiar momentum-winding picture of the compact boson is reorganised on the lattice into the bundle-flux data appropriate to fibred T-duality.

%%%%%%%%%%%%%%%%%%%%%%%%%%%%%%%%%%%%%%%%%%%%%%
%%%%%%%%%%%%%%%%%%%%%%%%%%%%%%%%%%%%%%%%%%%%%%
\subsection{Lightning review of the (modified) Villain model}
\label{eq:reviewmodifiedVillain}

The modified Villain formulation of the two-dimensional Euclidean XY model provides a lattice realisation of the compact boson model where both the momentum and winding $U(1)$ symmetries are exact, instead of emerging in the infrared.  These exact $U(1)_m\times U(1)_w$ symmetry statements are a special feature of the modified Villain model. In the twisted Villain model introduced in section \ref{sec:twistedVillainmodel}, this condition is replaced by background bundle and flux data, and the corresponding symmetry structure is modified accordingly; we analyse this in detail in section \ref{sec:symm_tV}.

Let $\G$ be a two-dimensional square Euclidean lattice, on which fields are defined as cochains. We denote as $C^m(\G,\mc{K})$ a cochain of rank $m$ taking value in the set $\mc{K}$, and as $\Delta:C^m(\G,\mc{K})\to C^{m+1}(\G,\mc{K})$ the lattice exterior derivative. The cochain analog of the wedge product is the cup product $\cup :  C^m(\G,\mc{K}) \times  C^n(\G,\mc{K}) \to C^{m+n}(\G,\mc{K}) $. We collect basic facts about cochains and cup products, together with details on the lattice and cochain notation used throughout the paper, in appendix \ref{app:cochains}. In this paper, the set $\mc{K}$ will usually be either $\mathbb R$ or $\mathbb Z$. Indeed, in the Villain formulation \cite{villain1975theory}, instead of working with circle-valued cochains, one introduces a real-valued $0$-cochain $\phi \in C^0(\G,\mathbb R)$ and an integer-valued $1$-cochain $n  \in C^1(\G,\mathbb Z)$. 
At large coupling $R$, the XY model is approximated by the Villain action \cite{villain1975theory}
\begin{equation}\label{eq:villain_act}
S_{\rm V} = \frac{R^2}{4\pi}\sum_{\text{links}} \bigl(\Delta\phi -2\pi n \bigr)^2\ ,
\end{equation}
with periodic boundary conditions for $\phi$ and $n$. The compactness of the theory variables is implemented through the $\mathbb Z$ gauge redundancy
\begin{equation}\label{eq:villain_gauge}
\phi \sim \phi+2\pi k\ , \q \q n\sim n+\Delta k\ , \q \q  k\in C^0(\G,\mathbb Z)\ .
\end{equation}

The gauge-invariant field strength of $n$ is the plaquette $2$-cochain $\Delta n \in C^2(\G,\mathbb Z)$, which in the XY model is interpreted as the local vorticity. The modified Villain formulation is obtained by imposing the flatness constraint
\begin{equation}\label{eq:mV:null_vort}
\Delta n=0\ , 
\end{equation}
thereby suppressing vortices. To this purpose,  one introduces a Lagrange multiplier cochain $\widetilde\phi$ and writes the action\footnote{Throughout the paper, all cochains (including $\widetilde \phi$) are defined on the same lattice, without introducing a dual lattice. For a reference in which the modified Villain model is formulated in this way, see \cite{Jacobson:2024muj}.} \cite{Gorantla:2021svj}
\begin{equation}
\label{eq:modified_villain}
S_{\rm V} = \frac{R^2}{4\pi}\sum_{\text{links}} \bigl(\Delta\phi -2\pi n \bigr)^2 + i \sum_{{\rm plaq}} \D n \cup \widetilde\phi  \ .
\end{equation}
Since $n$ is an integer-valued cochain, the cochain $\widetilde\phi$ is characterised by the gauge redundancy
\begin{equation}
\label{eq:villain_gaugedual}
\widetilde \phi \sim  \widetilde \phi+2\pi \widetilde k \ ,  \q \q \q \widetilde k\in C^0(\G,\mathbb Z)\ .
\end{equation}

The theory with action (\ref{eq:modified_villain}) is the modified Villain model. It differs from the ordinary Villain model by the additional flatness constraint, which makes it exhibit the main properties of the continuum compact boson model.  
Indeed, the model \eqref{eq:modified_villain} displays two exact continuous global symmetries, associated to the transformations
\be
\phi \ra \phi + \a  \ , \q \q \q \widetilde\phi \ra \widetilde \phi + \widetilde\a \ ,
\ee
where $\a$ and $\widetilde \a$ are constant 0-cochains taking values in $[0,2 \pi)$, due to the gauge redundancies (\ref{eq:villain_gauge}) and (\ref{eq:villain_gaugedual}). As a result, both symmetries are $U(1)$, and we refer to them as momentum $U(1)_{m}$ and winding $U(1)_{w}$ symmetries. Notice that the winding transformation with parameter $\widetilde\a$ is a symmetry thanks to the periodic boundary conditions of $n$, which are, in turn, tied to the absence of vortices $\D n =0$. Taking into account charge conjugation, 
the global symmetry group is
\begin{equation}
(U(1)_m\times U(1)_w)\rtimes \mathbb Z_2\ .
\end{equation}
One can show that the lattice model (\ref{eq:modified_villain}) also reproduces the mixed 't Hooft anomaly between the two $U(1)$ symmetries \cite{Gorantla:2021svj}. Moreover, and more importantly for our purposes, the modified Villain model is also characterised by a T-duality that acts as in the continuum as
\begin{equation}
U(1)_m \longleftrightarrow U(1)_w\ , \qquad R \longrightarrow 1/R\ .
\end{equation}
The duality can be derived using Poisson resummation, as we review in detail in appendix \ref{app:Tdualitycomputations}. We will use this technique to show T-duality for our twisted Villain model.

The importance of this short review for the results presented in this section is twofold. First, the modified Villain model provides the basic microscopic lattice realisation of a compact boson to be twisted below. Second, it makes clear that the exact $U(1)_w$ symmetry is tied to the vortex-free condition $\Delta n=0$. In the fibred construction introduced next, this condition will be replaced by a topologically twisted one, and the fate of the ordinary modified-Villain $U(1)_w$ must therefore be reconsidered.

%%%%%%%%%%%%%%%%%%%%%%%%%%%%%%%%%%%%%%%%%%%%%%
%%%%%%%%%%%%%%%%%%%%%%%%%%%%%%%%%%%%%%%%%%%%%%
\subsection{The twisted Villain model}
\label{sec:twistedVillainmodel}
As reviewed in section \ref{sec:reviewTduality}, T-duality can be more generally established for non-linear sigma models (\ref{eq:NSLMaction2}) in which the target space $E$ is a $U(1)$ bundle over a base manifold $M$, even when this bundle is topologically non-trivial. We aim to construct a lattice action that reproduces the part of the continuum non-linear sigma model action \eqref{eq:NSLMaction2} that is relevant for T-duality, namely the fibre sector and its coupling to the corresponding bundle and flux data. Throughout this section, we focus on the terms involving the fibre field $\Phi$, as they are the ones that are relevant for T-duality, and just formally assume that the base action $S_{{\rm bas}}$ is written in the modified Villain language as well. In section \ref{sec:explicitexamples}, we will construct concrete examples in which the base action $S_{{\rm bas}}$ is also specified explicitly.

We start discussing how to generalise the modified Villain model to include non-trivial fibrations. We then introduce the lattice Wess-Zumino term and finally discuss the complete twisted model, with a particular attention to its global symmetries.

%%%%%%%%%%%%%%%%%%%%%%%%%%%%%%%%%%%%%%%%%%%%%%
\subsubsection{Non-trivial fibrations on the lattice}
The non-linear sigma model (\ref{eq:NLSMaction}) involves a set of fields $X^{i}$. In the lattice Villain formulation, we introduce for each of them a real-valued 0-cochain $x^{i}$ and, for those that are compact in the continuum, an integer-valued 1-cochain $n^{(x^{i})}$ and a dual real-valued 0-cochain $\widetilde x^{i}$ that dynamically suppresses the corresponding vortices.

Recalling the decomposition $X^{i} = (Y^{\mu}, \Phi)$, where $Y^{\mu}$ are base fields and $\Phi$ is the fibre field, we aim to construct a lattice version of the action (\ref{eq:NSLMaction2}). In the continuum, the one-form $A$ is a basic form, and is thus constructed from the fields $Y^{\mu}$. On the lattice, we therefore introduce a real-valued 1-cochain $a$ and an integer-valued 2-cochain $N$, which in explicit models arise from the lattice realisation of the base sector and encode the basic connection and its topological data. They combine to give the field strength according to
\be
\label{eq:deflatticeF}
f = \D a + 2 \pi N \ .
\ee
The 2-cochain $N$ (and hence $f$) is closed once the dual cochains $\widetilde y^{\mu}$ have been integrated out, i.e.\ when the constraint $\D n^{(y^{i})} = 0$ is imposed. 

The lattice analogues of the fibre fields are $\phi \in C^0(\G,\mathbb R)$ and $n \equiv n^{(\phi)}  \in C^1(\G,\mathbb Z)$. Recalling (\ref{eq:decompmetric}), from the cochains introduced so far, we define the combination
\be
\label{eq:deflatticexi}
\mc{X} = \D \phi - a - 2 \pi n \ . 
\ee
We want its discrete exterior derivative to give the curvature $f$. From (\ref{eq:deflatticeF}) and (\ref{eq:deflatticexi}), we therefore deduce that we need the following condition,
\be
\D n =  N \ .
\ee
Following the idea of the modified Villain model, we impose this constraint dynamically by introducing a 0-cochain $\widetilde\phi$. At this stage, it is convenient to isolate the fibre-sector part of the lattice action, since this is the piece that will play the central role in the twisted Villain construction and can be defined independently of whether an explicit lattice realisation of the base sector is available. We therefore define
\begin{equation} \label{eq:emV_action}
S_{\rm fib} = \frac{R^2}{4\pi}\sum_{\text{links}} \bigl(\Delta \phi - a - 2\pi n\bigr)^2 + i\sum_{{\rm plaq}} \bigl(\Delta n - N\bigr)\cup \widetilde\phi \ .
\end{equation}
When considering the full non-linear sigma model, the total action is
\begin{equation} \label{eq:emV_action_total}
S_{\rm tot}=S_{{\rm bas}}+S_{\rm fib}\ ,
\end{equation}
where $S_{{\rm bas}}$ governs the dynamics of the cochains $y^\mu$, from which the background data $a$ and $N$ are (possibly only formally) constructed. In the following, however, we will work primarily with the fibre-sector action \eqref{eq:emV_action}, which already contains the twisted lattice implementation of the non-trivial fibration.

The fibre-sector action \eqref{eq:emV_action} retains the Villain gauge redundancies \eqref{eq:villain_gauge} and \eqref{eq:villain_gaugedual}. When a full lattice realisation of the base sector is specified, there may in addition be gauge redundancies acting on the cochains $y^\mu$. In section \ref{sec:explicitexamples}, we provide concrete examples where such additional gauge transformations are made explicit.

A prototypical example of a theory in the class (\ref{eq:emV_action}), where the target space is a non-trivial $U(1)$ bundle, is the $S^{3}$ non-linear sigma model with no $H$-flux. It is known to be T-dual to a nonlinear sigma model with trivially fibred target space $S^{1} \times S^{2}$ and one unit of $H$-flux. We now turn to this class of models.

\begin{figure}[t]
    \centering
    \includegraphics[width=0.835\linewidth]{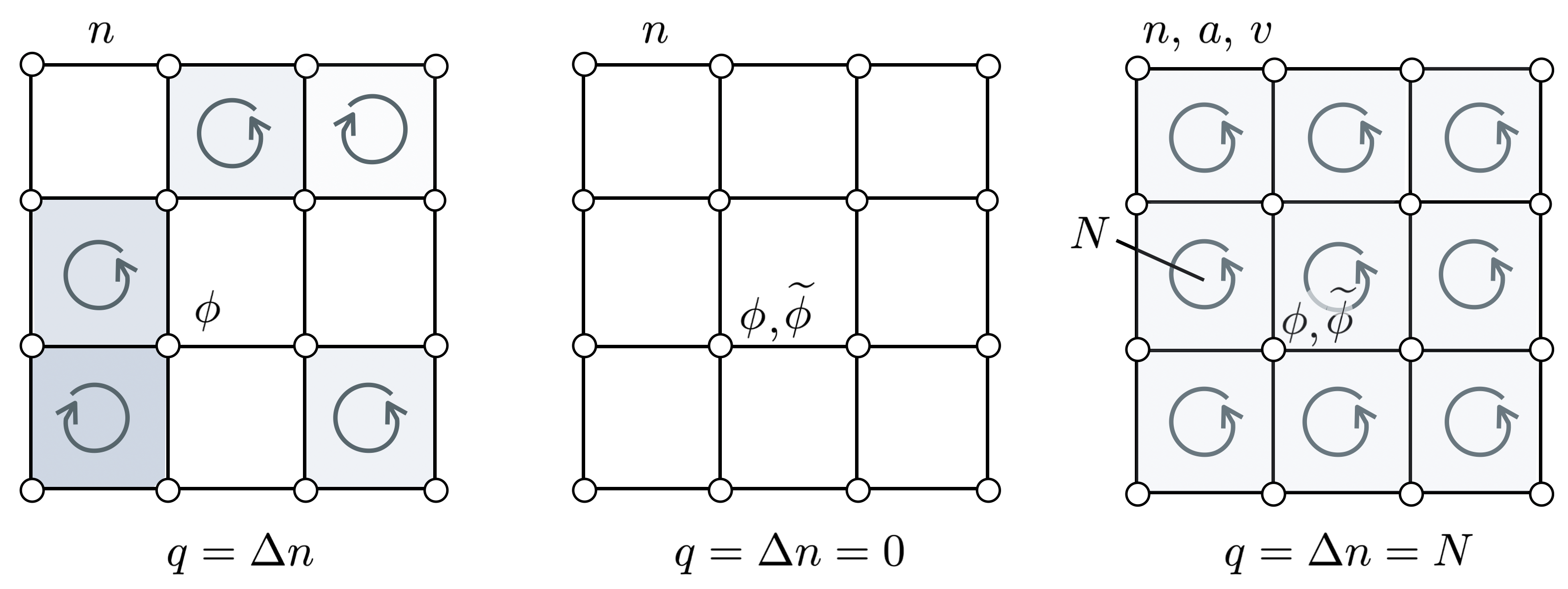}
    \caption{Comparative illustration of the ordinary, modified and twisted Villain constraints. Left: in the Villain model, plaquette vorticity is measured by the integer curl $q=\Delta n$. Middle: in the modified Villain model, flatness imposes $q=\Delta n=0$, suppressing vortices and yielding the exact dual-shift symmetry. Right: in the twisted model, the vortex constraint is shifted to $q=\Delta n=N$ by the  cocycle $N$, so vorticity is fixed by the bundle data rather than set to zero.}
    \label{fig:illustr_lattices_models}
\end{figure}

%%%%%%%%%%%%%%%%%%%%%%%%%%%%%%%%%%%%%%%%%
\subsubsection{Trivial fibration and non-trivial $H$-flux}\label{sec:triv_fib}

We aim to construct the lattice analogue of the last term in (\ref{eq:action_base_fiber}), namely the fibre-horizontal part of the Wess-Zumino term. Recalling the discussion in section \ref{seC:nonlinearsigmamodels}, we need to introduce the lattice counterpart of the one-form $V$, which was defined in \eqref{eq:iotaH_v}. Similarly to what we explained above for the one-form $A$, in the lattice case we therefore introduce a real-valued 1-cochain $v$, formally constructed from the base cochains $y^{\mu}$ and $n^{(y^{\mu})}$, as well as an integer-valued 2-cochain $\widetilde N$ constructed from $n^{(y^{\mu})}$. The associated field strength reads 
\be
\label{eq:deflatticetildeF}
\widetilde f = \D v + 2 \pi \widetilde N \ .
\ee
Integrating out the dual base cochains $\widetilde y^\mu$ imposes the flatness conditions on the base integer cochains, from which $\Delta \widetilde N=0$ follows. In that case, $\widetilde f$ is closed as well. In section \ref{sec:explicitexamples}, we will see this in an explicit example. 

For the present case of trivial fibration and non-trivial $H$-flux, we define the fibre action
\ba
\label{eq:emV_actionHflux} 
S_{{\rm fib}} &=&   \frac{R^2}{4\pi}\sum_{\text{links}} \pr{\D \phi  - 2 \pi n}^2  + i \sum_{{\rm plaq}}  \D n  \cup  \widetilde\phi 
+ i \sum_{\text{plaq}} \pq{\frac{1}{2 \pi} \pr{ \D \phi - 2 \pi n  } \cup  v  - \phi \cup \tilde N}  \ . \q \q 
\ea
The last term defines the lattice analogue of $B_{{\rm fib}}$ defined in (\ref{eq:B_decomp}), that is
\be
b_{{\rm fib}} = - \frac{1}{2 \pi} \pr{ \D \phi - 2 \pi n  } \cup  v  + \phi \cup \tilde N \ . \label{eq:bfib}
\ee
Taking its discrete exterior derivative,
\ba
\D b_{{\rm fib}}
&=&  \frac{1}{2 \pi} \pr{ \D \phi - 2 \pi n  } \cup \D  v  + \D \phi \cup \tilde N \nb \\
&=&  \frac{1}{2 \pi} \pr{ \D \phi - 2 \pi n  } \cup \pr{\D  v + 2 \pi \widetilde N}  + 2 \pi n   \cup   \widetilde N  \ ,  
\ea
where we used that $n$ and $\widetilde N$ are closed inside the partition function. More precisely, the integration over the fibre Lagrange multiplier $\widetilde\phi$ in the action \eqref{eq:emV_actionHflux} imposes $\Delta n=0$, while $\D N=0$ follows from the flatness conditions on the base integer cochains, as explained above. As a result, defining 
\be
h_{{\rm fib}} = \D b_{{\rm fib}} + 2 \pi N^{(\text{flux})} \ , \q \q \q N^{(\text{flux})} = -  n   \cup   \widetilde N \ , 
\ee
we have
\be
\label{eq:latticeHcochain}
h_{{\rm fib}} =  \frac{1}{2 \pi} \pr{ \D \phi - 2 \pi n  } \cup \widetilde f \ .
\ee
This indeed matches the continuum expression (\ref{eq:H_decomp2}). 

Actions terms such as the Wess-Zumino terms are notoriously subtle. While its proper definition at the lattice level may warrant a systematic study akin to that performed in \cite{Jacobson:2023cmr} for the lattice Chern-Simons action, the action (\ref{eq:emV_actionHflux}) captures the aspects of the Wess-Zumino coupling discussed in section \ref{sec:reviewTduality} that are relevant for fibrewise T-duality. Indeed, as we detail in appendix \ref{app:Tdualitycomputations}, using Poisson resummation, one can rewrite the theory (\ref{eq:emV_actionHflux}) as 
\be
\label{eq:emV_action2}
\widetilde S_{{\rm fib}} =  \frac{1}{4\pi R^2}\sum_{\text{links}} \pr{\D \widetilde \phi - v - 2 \pi \widetilde n}^2 + i \sum_{{\rm plaq}}   \phi \cup \pr{\D \widetilde n - \widetilde N}     \ .
\ee
This action is of the form (\ref{eq:emV_action}).  We therefore see that the cochains $v$ and $\widetilde N$ encoding the $H$-flux, after Poission resummation, become cochains that encode the fibration of the non-linear sigma model's target space.  
This is the behaviour expected by T-duality in non-linear sigma models and reviewed in section \ref{sec:bundle-flux-exchange}. 

%%%%%%%%%%%%%%%%%%%%%%%%%%%%%%%%%%
\subsubsection{Non-trivial fibration and non-trivial $H$-flux}
We now have all the ingredients needed to consider the general case in which the target space is non-trivially fibred and $H$-flux is present: the lattice analogues of the fibred kinetic term \eqref{eq:emV_action} and the fibre-horizontal coupling \eqref{eq:emV_actionHflux}. Combining the elements, we define the fibre part of the complete twisted Villain model as
\ba
\label{eq:emV_actioncomplete} 
S_{{\rm fib}} &=&   \frac{R^2}{4\pi}\sum_{\text{links}} \pr{\D \phi -a  - 2 \pi n}^2  + i \sum_{{\rm plaq}}  \pr{ \D n - N}  \cup  \widetilde\phi \nb \\
&+& i \sum_{\text{plaq}} \pq{\frac{1}{2 \pi} \pr{ \D \phi - a - 2 \pi n  } \cup  v  - \phi \cup \tilde N}  \ ,
\ea
which is the elementary lattice action needed to capture fibred T-duality at finite lattice spacing. Indeed, as detailed in appendix \ref{app:Tdualitycomputations}, using Poisson resummation, we obtain a dual model in terms of a fibre action
\begin{align} \label{eq:emV_actioncompletedual}
\widetilde  S_{{\rm fib}} &=   \frac{1}{4\pi R^2}\sum_{\text{links}} \left(\Delta \widetilde{\phi} - v - 2\pi \widetilde{n}\right)^2 + i \sum_{{\rm plaq}} \phi \cup \left(\Delta \widetilde{n} - \widetilde{N}\right) \nonumber\\
&\quad - i \sum_{\text{plaq}} \left[ \frac{1}{2\pi} a \cup \left(\Delta \widetilde{\phi} - 2\pi \widetilde{n}\right) + N \cup \widetilde{\phi} \right]\ .
\end{align}

We therefore see that,   upon Poisson resumming the twisted Villain model, we obtain the exchange 
\begin{equation}
a \longleftrightarrow v\ , \q \q \q  N \longleftrightarrow \widetilde{N}\ .
\end{equation}
This realises the lattice counterpart of the characteristic flux-bundle exchange of fibred T-duality  reviewed in section \ref{sec:reviewTduality}, and is one of the central results of our construction. In particular, it shows that the expected exchange of topological data admits an exact cochain realisation at finite lattice spacing. 

Let us conclude this section by noticing that in the complete model \eqref{eq:emV_actioncomplete}, the identification of the $h_{{\rm fib}}$ cochain is not as straightforward as in \eqref{eq:latticeHcochain}. Indeed, in the presence of $N$, the condition $\Delta n = N$ enforces twisted boundary conditions on $n$. In order to make the right identification, we define new cochains $n'$ and $a'$ such that
\begin{equation}
a + 2\pi n = a' + 2\pi n'\ , \qquad \Delta n' = 0\ , \qquad \Delta a' = f\ .
\end{equation}
The cochain $n'$ satisfies periodic conditions, while $a'$ does not.
We will then identify
\begin{equation} \label{eq:identifyH}
h_{{\rm fib}} = \Delta b_{{\rm fib}} + 2\pi N^{(\text{flux})}\ , \qquad N^{(\text{flux})} = - n' \cup \widetilde{N}\ ,
\end{equation}
which satisfies $\Delta h_{{\rm fib}} = 0$. In terms of the cochains $a'$ and $n'$, the $b_{{\rm fib}}$ cochain reads
\begin{equation}
b_{{\rm fib}} = - \frac{1}{2\pi} \left(\Delta \phi - a' - 2\pi n'\right)\cup v + \phi \cup \widetilde{N}\ .
\end{equation}
Its lattice exterior derivative yields
\begin{equation}
\Delta b_{{\rm fib}}=\frac{1}{2\pi}\left(\Delta \phi - 2\pi n'\right)\cup \left(\Delta v + 2\pi \widetilde{N}\right)+ 2\pi n' \cup \widetilde{N} + \frac{1}{2\pi}\Delta \left(a' \cup v\right)\ .
\end{equation}
The last term is basic and closed, and therefore can be included in the base part of the complete $h$ cochain,
\begin{equation}
\label{eq:Hidentificationgeneral}
h= h_{{\rm bas}}+ h_{{\rm fib}} \ , \q \q \q h_{{\rm fib}} =  \frac{1}{2\pi}\left(\Delta \phi - 2\pi n'\right)\cup \left(\Delta v + 2\pi \widetilde{N}\right)\ . 
\end{equation}
This matches the decomposition \eqref{eq:H_decomp2}.

%%%%%%%%%%%%%%%%%%%%%%%%%%%%%%%%%%%%%%%%%%%%%%
%%%%%%%%%%%%%%%%%%%%%%%%%%%%%%%%%%%%%%%%%%%%%%
\subsection{Global symmetries of the twisted Villain model}\label{sec:symm_tV}
We now analyse the global shift symmetries of the twisted Villain model. Since the shift transformations do not affect the base-dependent part of the action, we can restrict our attention to the fibre-dependent term \eqref{eq:emV_actioncomplete}. We determine the corresponding exact lattice symmetries by promoting constant shift parameters to local $0$-cochains, following the Noether procedure.

First, consider the shift of the fibre field, $\phi\longrightarrow \phi+\alpha$, with $\alpha$ a real $0$-cochain. Denoting as before $\mc{X} \equiv \Delta\phi-a-2\pi n$, the variation of the action \eqref{eq:emV_actioncomplete} is
\begin{equation}
\delta_\alpha S_{\rm fib} = \frac{R^2}{2\pi}\sum_{\ell:{\rm link}} \mc{X}_\ell(\Delta\alpha)_{\ell} +\frac{i}{2\pi}\sum_{\rm plaq}(\Delta\alpha)\cup v -i\sum_{\rm plaq}\alpha\cup\widetilde N\ .
\end{equation}
Using the cubical cup-product convention (see appendix \ref{app:cochains} for details), one has on a square lattice
\begin{equation}
\sum_{\rm plaq} (\Delta\alpha)\cup v= \sum_x\Big[(\Delta\alpha)_{x,1}\,v_{x+\hat1,2}-(\Delta\alpha)_{x,2}\,v_{x+\hat2,1}\Big] = \sum_{\ell:{\rm link}} \mathcal V_\ell(\Delta\alpha)_\ell\ ,
\end{equation}
where the last equality simply defines the induced link-cochain $\mathcal V$ by collecting the coefficients of $(\Delta\alpha)_\ell$.  We may therefore write the variation as
\begin{equation}
\delta_\alpha S_{\rm fib} = -i\sum_{\rm link} J_{{\rm iso},\ell}(\Delta\alpha)_\ell -i\sum_{\rm plaq}\alpha\cup \widetilde N \ , \qquad J_{\rm iso}=i\frac{R^2}{2\pi}\mc{X}-\frac{1}{2\pi}\mathcal V\ .
\end{equation}
On a square lattice with cochains satisfying periodic boundary conditions, using that $\sum_{\rm plaq} \alpha\cup \widetilde N= \sum_x \alpha_x\,\widetilde N_{x,12}$ and summing by parts, we obtain
\begin{equation}
\delta_\alpha S_{\rm fib} = i\sum_x \alpha_x\left(\Delta_\mu J_{{\rm iso},\mu}-\widetilde N_{x,12}\right)\ .
\end{equation}
Hence, the local Ward identity is
\begin{equation}
\Delta_\mu J_{{\rm iso},\mu}=\widetilde N_{x,12}\ .
\end{equation}
In the untwisted limit $v=\widetilde N=0$, this reduces to the usual modified-Villain momentum current and its conservation law.  

For a constant shift $\alpha$, the derivative terms vanish, and since a constant $0$-cochain factors out of the cup product, one finds
\begin{equation}
\delta_\alpha S_{\rm fib}=-\,i\alpha\sum_{\rm plaq} \widetilde N\ .
\end{equation}
This variation would vanish if $\widetilde N$ were an exact cochain, namely if it were possible to find a 1-cochain $r$ satisfying periodic boundary conditions and such that $\widetilde N = \D r$. However, depending on the case at hand, this need not hold in general. Indeed, there are cases in which $\widetilde N$ is not exact.\footnote{See \eqref{eq:latticeT3v} for a concrete example.} In such situations, the action is not invariant under generic constant continuous shifts $\a$.
Recalling that $\widetilde N$ is the cochain that encodes the (pullback on the lattice of the) dual bundle's Chern-class, in general, we have\footnote{If the case at hand corresponds to one in which the target-space base $M$ admits several 2-cycles, the most general expression is, in fact, 
\[
\sum_{\rm plaq} \widetilde N \in \sum_i \kappa_i \, \mathbb{Z}\ .
\qquad \kappa_i \in \mathbb{N}\ .
\]
In this case, the discussion proceeds in the same way after defining $\kappa$ as the  greatest common divisor of the $\kappa_i$. An analogous comment applies to $\lambda$ below.}
\be
\sum_{\rm plaq} \widetilde N \in \kappa \, \mathbb Z \ , \q \q \q \kappa \in\mathbb N \ .
\ee
Then, the exact symmetry of the lattice path-integral weight is the subgroup satisfying
$e^{i\alpha \kappa \mathbb Z}=1$, namely
\begin{equation}
U(1)_{\text{iso}} \longrightarrow \mathbb Z_{ \kappa}\ ,
\end{equation}
with the understanding that $ \kappa=0$ restores the full $U(1)$.

We now turn to the dual shift, $\widetilde\phi\longrightarrow \widetilde\phi+\widetilde\alpha$. Only the Lagrange multiplier term varies:
\begin{equation}
\label{eq:stepcalculationsymmetry}
\delta_{\widetilde\alpha}S_{\rm fib}=i\sum_{\rm plaq}(\Delta n-N)\cup\widetilde\alpha\ .
\end{equation}
For constant $\widetilde\alpha$, using that $n$ satisfies periodic boundary conditions off-shell, this becomes
\begin{equation}
\delta_{\widetilde\alpha}S_{\rm fib}=-\,i\widetilde\alpha\sum_{\rm plaq} N\ .
\end{equation}
Analogously to the discussion above for $\widetilde N$, the cochain $N$ is not necessarily exact. In general we have\footnote{See (\ref{eq:latticeAnilfold}) for a concrete example.}
\be
\sum_{\rm plaq}  N \in  \l \, \mathbb Z \ , \q \q \q \l \in\mathbb N \ .
\ee
As a result, the dual shift transformations are symmetries provided $e^{i\tilde \alpha \l \mathbb Z}=1$, hence the dual isometry symmetry will be broken to a discrete subgroup, 
\begin{equation}
U(1)_{\widetilde{\text{iso}}} \longrightarrow \mathbb Z_{\l}\ ,
\end{equation}
with again the understanding that $\l=0$ restores the full $U(1)$. For $\l=0$, the model features a gauge-invariant conserved current associated to the $U(1)_{\tilde \phi}$ symmetry,
\begin{equation}
J_{\widetilde{\text{iso}},\mu}=\frac{\varepsilon_{\mu\nu}}{2\pi}\left(\Delta_\nu\phi-2\pi n_\nu\right)\ ,  \q \q \q
\Delta_\mu J_{\widetilde{\text{iso}},\mu} = 0 \ , \q \q \q \text{for $ \ \l=0$} \ .
\end{equation}
For $\l \neq 0$, this current is not conserved anymore, 
\begin{equation}
\Delta_\mu J _{\widetilde{\text{iso}},\mu} =-\,\Delta n=-\,N\ .
\end{equation}

Below we contrast and summarise these symmetry patterns in turn with the modified Villain model and fibred non-linear sigma-models, see also table \ref{tab:symmetry_comparison} for a succinct summary.
\begin{table}[t]
\centering
\renewcommand{\arraystretch}{1.25}
\setlength{\tabcolsep}{2pt}
\begin{tabularx}{\textwidth}{@{}l>{\centering\arraybackslash}X>{\centering\arraybackslash}X@{}}
\textbf{Theory} & \textbf{Geometric input} & \textbf{Symmetry pattern} \\
\hline
modified Villain
&
$\Delta n=0$
&
$U(1)_m \times U(1)_w$
\\[1mm]
\hline
\\[-5mm]
twisted Villain
&
\(
\Delta n = N
\qquad
\begin{gathered}
\sum_{\rm plaq} \smt N \in \kappa \mathbb Z \\
\sum_{\rm plaq} N \in \lambda \mathbb Z
\end{gathered}
\)
&
\(
\begin{gathered}
U(1)_{\rm iso}\to \mathbb Z_\kappa \\
U(1)_{\smt{\rm iso}}\to \mathbb Z_{\lambda}
\end{gathered}
\)
\\[4mm]
\hline
\\[-5mm]
fibred NLSM
&
\(
\begin{gathered}
\smt F \text{ non-exact on } E \\
F \text{ non-exact on } \smt E
\end{gathered}
\)
&
\(
\begin{gathered}
U(1)_{\rm iso}\to \mathbb Z_\kappa \\
U(1)_{\smt{\rm iso}}\to \mathbb Z_{\lambda}
\end{gathered}
\)
\\
\end{tabularx}
\caption{Comparison of the symmetry structure of the modified Villain model, the twisted Villain model, and the continuum fibred non-linear sigma model.}
\label{tab:symmetry_comparison}
\end{table}

%%%%%%%%%%%%%%
\paragraph{Comparison with the modified Villain model.}

As reviewed in section \ref{eq:reviewmodifiedVillain}, the modified Villain model is based on the flatness condition
\begin{equation}
\Delta n=0\ ,
\end{equation}
which suppresses vortices and yields two exact continuous symmetries, acting on $\phi$ and $\widetilde\phi$, respectively. This flatness condition lies at the origin of the exact $U(1)_m\times U(1)_w$ structure and of the corresponding lattice T-duality reviewed in section~\ref{eq:reviewmodifiedVillain}. In the twisted Villain model, by contrast, flatness is replaced by the topological constraint
\begin{equation}
\Delta n=N  \ , 
\end{equation}
so that vorticity is no longer suppressed but fixed by the background bundle data. As a result, the exact continuous symmetries of the modified Villain model are generically reduced, and the natural lattice T-duality is no longer organised by a pure exchange of momentum and winding. Instead, the relevant input at the level of this lattice realisation is the bundle-flux data encoded by $N$ and $\widetilde N$. The twisted Villain model however does realise T-duality through the general form appropriate to fibred curved backgrounds, where the duality exchanges topological and flux data rather than only the momentum-winding charges of the free boson.

%%%%%%%%%%%%%%
\paragraph{Comparison with continuum fibred sigma models.} 
The relation to fibred non-linear sigma models is much closer, but conceptually different in a crucial aspect. In the ordinary compact-boson case, the isometric circle, the existence of a genuine winding sector, and the second exact $U(1)$ symmetry are naturally identified. In fibred sigma models, however, these notions need no longer coincide. As emphasised in \cite{Arias-Tamargo:2025xdd}, the relevant pair of group-like symmetries is instead the isometry and dual-isometry symmetry. The first is associated with shifts along the fibre Killing vector and is controlled by whether $\iota_k H$, or equivalently the corresponding $\widetilde F$, is globally exact; the second is most naturally described in the T-dual frame and is controlled analogously by the topology encoded in $F$. In particular, the ordinary free-boson winding symmetry need not survive as an independent global symmetry of the full target-space sigma model.

The relation to the lattice is therefore qualitatively close, but differs in how the relevant topology is encoded. In the continuum, the bundle curvatures $F$ and $\widetilde F$ are intrinsically determined by the target-space geometry, and the corresponding isometry and dual-isometry symmetries are controlled by whether these curvature are globally exact or not. On the lattice, the analogous information is encoded through the cochains $a,v,N,\widetilde N$ and the twisted vorticity constraint $\Delta n=N$, which determine the cochain field strength $f$ and $\widetilde f$. In this sense, the same symmetry pattern is organised by the global exactness properties of the relevant bundle and flux data on the lattice as well, even though these data are implemented through vorticity rather than inherited from an underlying smooth geometry.

Despite this difference, the resulting symmetry pattern is qualitatively the same as in the continuum fibred models \cite{Arias-Tamargo:2025xdd}. The ordinary free-boson interpretation in terms of exact momentum and winding symmetries is replaced by the more appropriate pair of fibre isometry and dual isometry, whose precise form is controlled by the bundle-flux data. The Hopf fibration $S^1\hookrightarrow S^3\to S^2$ is the basic example: the fibre remains an isometric circle along which T-duality can be performed, but it does not define an independent free-boson winding sector. In this respect, the twisted Villain model reproduces the symmetry patterns of the non-linear sigma models: what survives in curved fibred backgrounds is not an exact momentum-winding interpretation, but the symmetry structure dictated by bundle-flux exchange.

%%%%%%%%%%%%%%%%%%%%%%%%%%%%%%%%%%%%%%%%%%%%%%
%%%%%%%%%%%%        Examples       %%%%%%%%%%%
%%%%%%%%%%%%%%%%%%%%%%%%%%%%%%%%%%%%%%%%%%%%%%
\section{Lattice realisations of fibred string backgrounds}
\label{sec:explicitexamples}
In section \ref{sec:twistedVillain}, we presented the twisted Villain model and its T-duality properties in a general setting, where the cochains $a$, $v$, $N$, and $\widetilde N$ are assumed to depend on the base cochains. In the present section, we show how this general construction is realised in explicit lattice versions of well-known backgrounds arising in string compactifications. In these concrete examples, we verify that the twisted Villain model reproduces the expected bundle and flux data, their associated symmetry pattern, and the characteristic bundle-flux exchange under T-duality.

The first three are cases in which the base sector can itself be realised within the lattice framework: they are all $U(1)$ bundles over a two-torus $T^2$, namely the trivially fibred $T^3$ model with zero $H$-flux, the corresponding model with non-zero $H$-flux, and the nilfold. We then turn to the Hopf fibration, which highlights a different aspect of the construction: although the curved base is not explicitly discretised, the twisted Villain model still realises the fibre-sector topological data that control fibred T-duality.

%%%%%%%%%%%%%%%%%%%%%%%%%%%%%%%%%%%%%%%%%%%%%%
%%%%%%%%%%%%%%%%%%%%%%%%%%%%%%%%%%%%%%%%%%%%%%
\subsection{Three-torus model with zero $H$-flux}
The sigma model with target space $E = T^{3}$ and no $H$-flux is equivalent to three copies of the compact boson model. We present it here mostly to set the notation for the next cases. The base is a two-torus $M=T^{2}$, and its coordinates $Y^{\m}=(\Theta, \Psi)$ are both taken to have period $2\pi$. The fibre $\Phi$ is, in this case, another independent circle with coordinate $\Phi \in [0,2 \pi)$. In the modified Villain model, we introduce the real-valued 0-cochains $\pg{\th, \psi, \phi}$, the integer-valued 1-cochains $\pg{n^{(\th)}, n^{(\psi)}, n}$, the real-valued 0-cochains $\pg{\widetilde \th,\widetilde  \psi, \widetilde  \phi}$.
and write the action
\ba
S &=& \frac{R^{2}}{4 \pi} \sum_{\text{links}} \pq{ \pr{\D \theta - 2 \pi n^{(\th)}}^{2} + \pr{\D \psi - 2 \pi n^{(\psi)}}^{2} 
+ \pr{\D \phi - 2 \pi n}^{2}} \nb \\
&+& i \sum_{{\rm plaq}} \pr{  \D n^{(\th)} \cup \tilde \theta + \D n^{(\psi)} \cup \tilde \psi  +  \D n \cup \ \tilde \phi } \ .
\ea
The action is invariant under the gauge transformations
\begin{subequations}
\label{eq:torusVillaingauge}
\begin{align}
\phi &\ra \phi+2\pi k  \ , 
& n &\ra n+\Delta k  \ , 
& \tilde\phi &\ra \tilde\phi+2\pi \tilde k \ ,  \\
\label{eq:torusVillaingaugetheta}
\theta &\ra \theta + 2\pi k^{(\theta)} \ , 
& n^{(\theta)} &\ra n^{(\theta)}+\Delta k^{(\theta)}  \ , 
& \tilde\theta &\ra \tilde\theta+2\pi \tilde k^{(\theta)}  \ , \\
\label{eq:torusVillaingaugepsi}
\psi &\ra \psi+2\pi k^{(\psi)}  \ , 
& n^{(\psi)} &\ra n^{(\psi)}+\Delta k^{(\psi)}  \ , 
& \tilde\psi &\ra \tilde\psi+2\pi \tilde k^{(\psi)} \ ,
\end{align}
\end{subequations}
where all the $k$ are integer-valued 0-cochains.

\begin{figure}[t]
    \centering
    \includegraphics[width=0.38\linewidth]{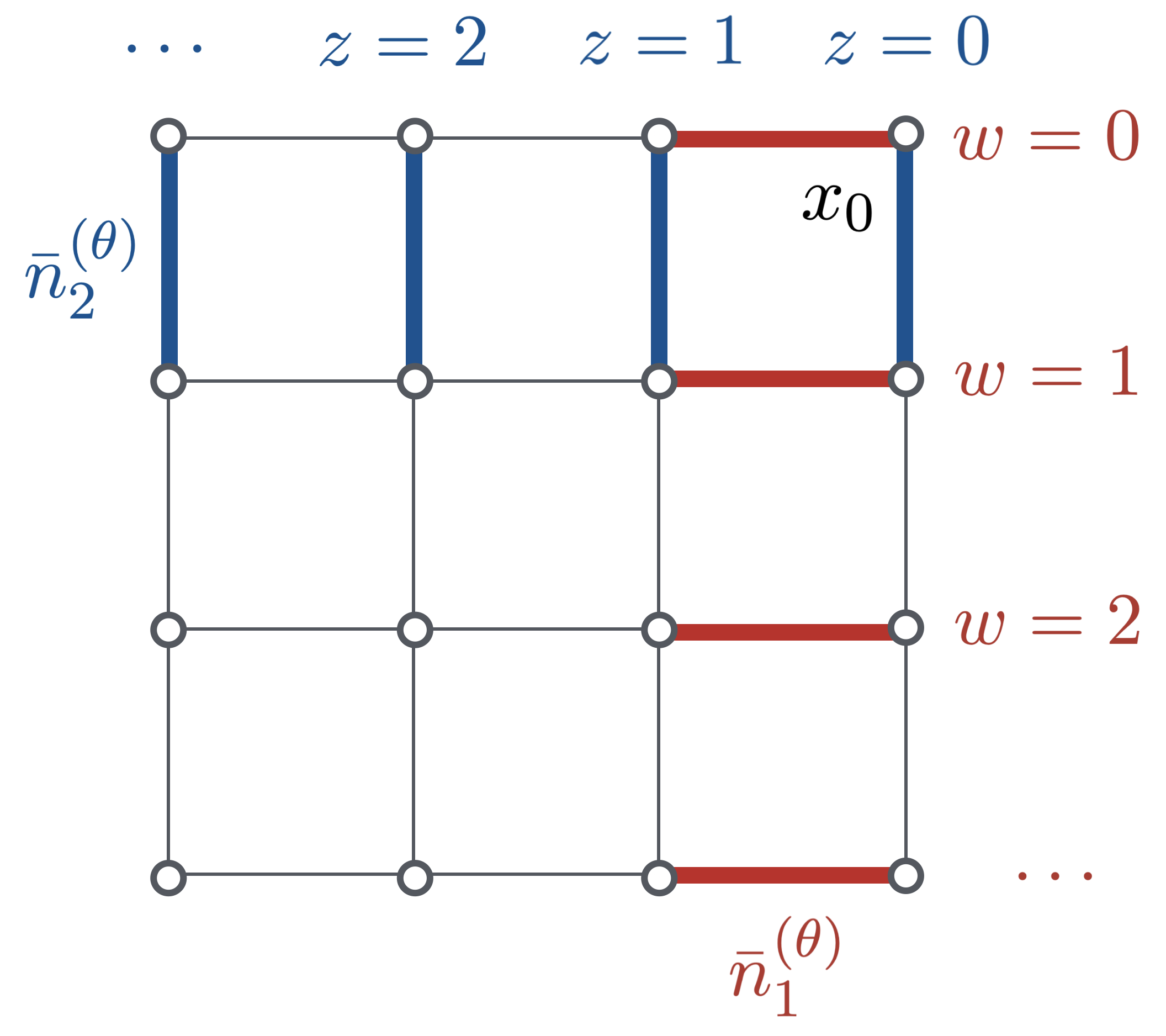}
    \caption{Gauge-fixed representative of the flat integer one-cochains on the periodic square lattice. For $x_0$ the upper-right site, all links are gauged away except the representatives in \eqref{eq:baseintegerfields}: the red horizontal links carry $\bar n^{(\theta)}_1$, while the blue vertical links carry $\bar n^{(\theta)}_2$. The same convention is used for $n^{(\psi)}$, with $\bar n^{(\theta)}_{1,2}$ replaced by $\bar n^{(\psi)}_{1,2}$.}
    \label{fig:illustr_gaugefix_torus}
\end{figure}

In the next two examples, the base sector will be a two-torus. It will therefore be given by two copies of the modified Villain model. To make the formulas transparent, it will be actually convenient to perform gauge fixing in the base sector. For this purpose, we follow appendix B.1.4 of \cite{Gorantla:2021svj}. The idea is to integrate out the Lagrange mutiplier cochains $\tilde \th$ and $\tilde \psi$ so that $\D n^{(\th)}= \D n^{(\psi)} = 0$ and then fix (\ref{eq:torusVillaingaugetheta}) and (\ref{eq:torusVillaingaugepsi}) so that $n^{(\th)}=0 $ and $n^{(\psi)}=0$ everywhere but for 
\begin{subequations}
\label{eq:baseintegerfields}
\begin{align}
&n^{(\th)}_{x_{0} -  \hat 1 - w \hat 2,1} = \bar n^{(\th)}_{1} \ , &w \in \pg{0,1,\dots, L_{2}-1} \ ,  \\
&n^{(\psi)}_{x_{0} -   \hat 1 - w \hat 2,1} = \bar n^{(\psi)}_{1} \ , &w \in \pg{0,1,\dots, L_{2}-1} \ ,  \\
&n^{(\th)}_{x_{0} - z \hat 1 -  \hat 2,2} = \bar n^{(\th)}_{2} \ , &z \in \pg{0,1,\dots, L_{1}-1} \ , \\
&n^{(\psi)}_{x_{0} - z \hat 1 - \hat 2,2} = \bar n^{(\psi)}_{2} \ , &z \in \pg{0,1,\dots, L_{1}-1} \ .
\end{align}
\end{subequations}
Here, we denoted with $x_{0}$ the right-upper lattice site and with $L_{1}$ and $L_{2}$ the number of lattice sites along the two directions. For a useful illustration of the gauge-fixed configuration, see figure \ref{fig:illustr_gaugefix_torus}. We are left with four integers, counting the winding of the two base fields along the two directions. 
After this gauge-fixing, there are residual gauge symmetries  \cite{Gorantla:2021svj}
\be
\label{eq:nilfoldbaseresidualgauge}
\th \sim \th + 2 \pi q^{(\th)} \ , \q \q \q \psi \sim \psi + 2 \pi q^{(\psi)} \ ,
\ee
where $q^{(\th)}$ and $q^{(\psi)}$ are constant integer-valued 0-cochains. 
The (base part of the) partition function will be given by the integral over the real-valued $\th$ and $\psi$ at any lattice site and the sum over the four integers (\ref{eq:baseintegerfields}).

%%%%%%%%%%%%%%%%%%%%%%%%%%%%%%%%%%%%%%%%%%%%%%
%%%%%%%%%%%%%%%%%%%%%%%%%%%%%%%%%%%%%%%%%%%%%%
\subsection{Three-torus model with $H$-flux}
We next consider the $T^3$ sigma-model in the presence of non-trivial $H$-flux. In the fibered description reviewed in section \ref{sec:reviewTduality}, the relevant component is the fibre contribution
\be
\label{eq:Hfluxdef}
\int_{T^{3}} H_{{\rm fib}}  = \frac{1}{(2 \pi)^{2}} \int_{T^{3}} \dd \Phi \we \dd \Theta \we \dd \Psi = 2 \pi  \kappa \ , \q \q \q \kappa \in \mathbb Z \ . 
\ee
Equation \eqref{eq:iotaH_v} then implies that the associated basic curvature is 
\be
\label{eq:tildeFHfluxT3}
\widetilde F = \frac{\kappa}{2 \pi}\dd \Theta \we \dd \Psi \ .
\ee
A potential that yields this field strength is
\be
\label{eq:vHfluxT3}
 V = \frac{\kappa}{2 \pi}  \Theta \dd \Psi \ , \q \q \q \widetilde F \equiv \dd V \ .
\ee
Notice that it is not possible to find a globally-defined one-form in $E=T^{3}$ that gives (\ref{eq:tildeFHfluxT3}). As a result, 
recalling the comments on the non-linear sigma model's global symmetries made in section \ref{sec:bundle-flux-exchange}, this is a case in which the isometry symmetry $\Phi \ra \Phi + \a$ with $\a$ constant is a $\mathbb Z_{\kappa}$ symmetry.

The lattice realisation of this background is obtained by introducing a 1-cochain $v$ as the lattice analogue of \eqref{eq:vHfluxT3}, together with an integer-valued 2-cochain $\widetilde N$ such that $\widetilde f=\Delta v+2\pi \widetilde N$ reproduces \eqref{eq:tildeFHfluxT3}. We define
\be
\label{eq:latticeT3v}
v = \frac{\kappa}{2 \pi} \th \cup \pr{ \D \psi - 2 \pi n^{(\psi)}} + \kappa n^{(\th)} \cup \psi \ , \q \q \q \widetilde N= \kappa n^{(\th)} \cup n^{(\psi)} \ .
\ee
Since $\D n^{(\th)}=\D n^{(\psi)}=0$, the cochain $\widetilde N$ is closed. It follows that
\be
\widetilde  f = \frac{\kappa}{2 \pi} \pr{ \D \th - 2 \pi n^{(\th)} }\cup \pr{ \D \psi - 2 \pi n^{(\psi)}} \ , \q \q \q \D \widetilde  f=0 \ .
\ee
Under the residual gauge symmetries \eqref{eq:nilfoldbaseresidualgauge} of the base sector, $v$ transforms as
\be
v \ra v + \kappa q^{(\th)} \pr{ \D \psi - 2 \pi n^{(\psi)}} + 2 \pi  \kappa  q^{(\psi)} n^{(\th)}  \ ,
\ee 
whereas $\D v$ and $\widetilde N$ are invariant. Consequently, the field strength $\widetilde f$ is gauge invariant as well.

The model's action is
\ba
\label{eq:T3latticeactionHflux}
S &=& \frac{R^{2}}{4 \pi} \sum_{\text{links}} \pq{ \pr{\D \th - 2 \pi n^{(\th)}}^{2} + \pr{\D \psi - 2 \pi n^{(\psi)}}^{2} 
+  \pr{\D  \phi  - 2 \pi   n}^{2}} \nb \\
&+&  i \sum_{{\rm plaq}}  \D n  \cup  \widetilde\phi  + i \sum_{\text{links}} \pq{\frac{1}{2 \pi} \pr{ \D \phi - 2 \pi n  } \cup  v  - \phi \cup \widetilde N}  \ .
\ea
The model displays the following gauge redundancy,
\begin{subequations}
\label{eq:T3withHfluxgaugetransf}
\ba
\th &\ra & \th + 2 \pi q^{(\th)} \ , \\
\psi &\ra & \psi + 2 \pi q^{(\psi)} \ , \\
\phi &\ra&  \phi + 2 \pi  k \ , \\
n &\ra&    n   + \D   k \ , \\
\widetilde \phi &\ra&  \widetilde \phi + \kappa q^{(\th)} \psi + 2 \pi \widetilde k \ ,
\ea
\end{subequations}
where $k$ and $ \widetilde k$ are integer-valued cochains. 
From the last term in (\ref{eq:T3latticeactionHflux}), we read the $b_{{\rm fib}}$ cochain
\ba
\label{eq:explicitbinT3model}
b_{{\rm fib}} &=& - \frac{\kappa}{(2 \pi)^{2}} \pr{ \D \phi - 2 \pi n  } \cup  \pq{\frac{1}{2 \pi} \th \cup \pr{ \D \psi - 2 \pi n^{(\psi)}} +  n^{(\th)} \cup \psi}  \nb \\
&+& \kappa \phi \cup \pr{n^{(\th)} \cup n^{(\psi)} } \ .
\ea
One can check that under the gauge transformations (\ref{eq:T3withHfluxgaugetransf}), the $b$ cochain (\ref{eq:explicitbinT3model}) transforms as 
\be
\label{eq:btransf}
b_{{\rm fib}} \ra b_{{\rm fib}} + \D \L + 2 \pi \ell  \ , 
\ee
where $\L$ is a real-valued 1-cochain and $\ell $ is an integer-valued 2-cochain that is closed  ($\D \ell=0$) once the condition $\D n =0$ is imposed.
The associated $h_{{\rm fib}}$ cochain is
\be
\label{eq:latticeHcochain2}
h_{{\rm fib}} =  \frac{\kappa}{(2 \pi)^{2}} \pr{ \D \phi - 2 \pi n  } \cup \pq{   \pr{ \D \th - 2 \pi n^{(\th)} }\cup \pr{ \D \psi - 2 \pi n^{(\psi)}} }   \ ,
\ee
which matches the continuum expression (\ref{eq:Hfluxdef}).

The symmetry pattern of this background is a concrete instance of the general discussion in section \ref{sec:symm_tV}. In the present case, $\widetilde N=\kappa\,n^{(\theta)}\cup n^{(\psi)}$, so the isometry shift $\phi\to\phi+\alpha$ varies the action as 
\begin{align}
	\delta S = - i \alpha \sum_{{\rm plaq}} \widetilde N = - i \kappa \alpha \sum_{{\rm plaq}} n^{(\theta)} \cup n^{(\psi)}  \in i \kappa \alpha \,\mathbb Z \ .
\end{align}
Hence, only the shifts with parameter $\alpha \in 2\pi \mathbb Z/\kappa$ leave the path-integral weight invariant, and the isometry symmetry is reduced to $\mathbb Z_\kappa$, exactly as predicted by the general analysis. By contrast, since in this example the fibration is trivial and therefore $N=0$, the dual shift $\widetilde\phi\to\widetilde\phi+\widetilde\alpha$ remains an exact continuous symmetry. The dual isometry symmetry is therefore $U(1)$.

This model is T-dual to the background with non-trivial $U(1)$ fibration over $M=T^2$ and vanishing $H$-flux, to which we now turn. 

%%%%%%%%%%%%%%%%%%%%%%%%%%%%%%%%%%%%%%%%%%%%%%
%%%%%%%%%%%%%%%%%%%%%%%%%%%%%%%%%%%%%%%%%%%%%%
\subsection{Nilfold model}
We now consider the simplest genuinely fibred example, namely the nilfold. This is a three-manifold realised as a non-trivial $U(1)$ bundle over $T^2$, and it is T-dual to the previous $T^3$ model with non-zero $H$-flux. It therefore provides the basic geometric setting in which the lattice twisted Villain model captures non-trivial bundle topology rather than only background flux.

The metric of the nilfold can be written as
\begin{align}
\label{eq:decompmetric2}
g = g_{\rm bas} + \frac{R^{2}}{2 \pi} \xi\otimes \xi \ , \q \q \q \xi \equiv \dd \Phi - A\ ,
\end{align}
where $g_{\rm bas}$ is the flat Euclidean metric and $A$ is a connection on the bundle $E$ with non-trivial Chern class,
\be
\int_{T^{2}}  F  = 2 \pi  \l \ , \q \q \q \l \in \mathbb Z \ ,  \q \q \q F \equiv \dd A \ .
\ee
We can choose
\be
\label{eq:AFnilfold}
A = \frac{\l}{2 \pi} \Theta \dd \Psi \ , \q \q \q F = \frac{\l}{2 \pi} \dd  \Theta\wedge \dd \Psi \ .
\ee
As in the previous example, there is no globally defined one-form on the nilfold whose exterior derivative gives \eqref{eq:AFnilfold}. Accordingly, the dual isometry shift $\widetilde\Phi \to \widetilde\Phi+\widetilde\alpha$ is reduced to a $\mathbb Z_\lambda$ symmetry.

The lattice realisation of the non-linear sigma model with nilfold target space is obtained by introducing a 1-cochain $a$ and an integer-valued 2-cochain $N$ as the lattice analogues of \eqref{eq:AFnilfold}. We define
\be
\label{eq:latticeAnilfold}
a = \frac{\lambda}{2 \pi} \th \cup \pr{ \D \psi - 2 \pi n^{(\psi)}} + \lambda n^{(\th)} \cup \psi \ , \q \q \q   N= \lambda n^{(\th)} \cup n^{(\psi)} \ .
\ee
We have $\D   N =0$ since $\D n^{(\th)}  = \D n^{(\psi)} =0$. We obtain
\be
  f = \frac{\lambda}{2 \pi} \pr{ \D \th - 2 \pi n^{(\th)} }\cup \pr{ \D \psi - 2 \pi n^{(\psi)}} \ , \q \q \q \D  f=0 \ .
\ee
The cochain $a$ transforms under the residual gauge symmetries (\ref{eq:nilfoldbaseresidualgauge}) of the base sector according to
\be
a \ra a + \lambda q^{(\th)} \pr{ \D \psi - 2 \pi n^{(\psi)}} + 2 \pi \lambda q^{(\psi)} n^{(\th)}  \ ,
\ee 
while $\D a$ and $ N$ are invariant. As a result, the field strength $f$ is invariant as well.
The model's action is
\ba
\label{eq:nilfoldaction}
S 
&=&\frac{R^{2}}{4 \pi} \sum_{{\rm plaq}} \pq{ \pr{\D \th - 2 \pi n^{(\th)}}^{2} + \pr{\D \psi - 2 \pi n^{(\psi)}}^{2} 
} \nb \\ 
&+& \frac{R^2}{4\pi}\sum_{{\rm plaq}} \pr{\D \phi - a - 2 \pi n}^2 + i \sum_{{\rm plaq}} \pr{\D n - N} \cup  \widetilde\phi  \ .
\ea
The model displays the following gauge redundancy,
\begin{subequations}
\label{eq:nilfoldgaugetransf}
\ba
\th &\ra & \th + 2 \pi q^{(\th)} \ , \\
\psi &\ra & \psi + 2 \pi q^{(\psi)} \ , \\
\phi &\ra& \phi + \lambda  q^{(\th)} \psi + 2 \pi k \ , \\
n &\ra& n - \lambda q^{(\psi)} n^{(\th)} + \lambda  q^{(\th)} n^{(\psi)}  + \D k \ ,\\
\widetilde \phi &\ra& \widetilde \phi + 2 \pi \widetilde k \ ,
\ea
\end{subequations}
where $k$ and $ \widetilde k$ are integer-valued cochains.

The symmetry pattern again matches the general discussion of section \ref{sec:symm_tV}. In the present case, the isometry shift $\phi \ra \phi+\alpha$ remains an exact $U(1)$ symmetry. By contrast, the dual isometry shift $\widetilde \phi \ra \widetilde \phi+\widetilde \alpha$ transforms the action as
\be
\d_{\widetilde \a} S = - i \widetilde \a \sum_{\text{links}}  N =  - i \lambda \widetilde \a \sum_{\text{links}}  n^{(\th)} \cup n^{(\psi)} \in - i \lambda \widetilde \a \mathbb Z \ .
\ee
Only the transformations with  $\widetilde \a \in 2 \pi \mathbb Z/\lambda$ are symmetries. Thus, the dual isometry symmetry is $\mathbb Z_{\lambda}$.

%%%%%%%%%%%%%%%%%%%%%%%%%%%%%%%%%%%%%%%%%%%%%%
%%%%%%%%%%%%%%%%%%%%%%%%%%%%%%%%%%%%%%%%%%%%%%
\subsection{Hopf-fibration model}
The previous examples presented in this section admit a fully explicit lattice realisation: the base sector can be modelled by modified Villain models, while the non-trivial fibration of the fibre is encoded by the twisted Villain model. In this way, one obtains a complete lattice model for the corresponding fibred sigma model. These examples, however, belong to a rather special class of backgrounds, in which the base itself can still be realised within the same lattice framework. More general curved geometries need not share this property. The simplest example is the Hopf fibration
\begin{equation}
S^1 \hookrightarrow S^3 \to S^2 \ .
\end{equation}
The full target-space theory is now the $SU(2)$ WZW model written in coordinates adapted to the circle fibre. Unlike the torus and nilfold cases, the base is $S^2$, and is therefore curved, while the target-space dynamics is genuinely non-linear. A lattice discretisation based solely on a combination of modified and twisted Villain models no longer provides a direct realisation of the full sigma model. For fibred T-duality, however, this is not required. What enters the duality is the fibre-adapted part of the action, reviewed in section \ref{seC:nonlinearsigmamodels},
\begin{equation}
S_{\rm fib}=\frac{R^2}{4\pi}\int_\Sigma (d \Phi-A)\wedge * (d \Phi-A) - \frac{i}{2\pi}\int_\Sigma V\wedge d \Phi\ ,
\end{equation}
together with the associated bundle and flux data $F=\mathrm dA$ and $\widetilde F=\mathrm d V$.  It is precisely this fibre-sector information that the twisted Villain model encodes. Accordingly, the Hopf-fibre sector is realised on the lattice by taking the twisted Villain model \eqref{eq:emV_actioncomplete} with fixed background cochains $a,v\in C^1(\G,\mathbb R)$ and $N\in C^2(\G,\mathbb Z)$ satisfying
\begin{equation}
f=\Delta a+2\pi N\ , \qquad \widetilde f=\Delta v\ .
\end{equation}
In the Hopf case, the non-trivial fibre topology is thus encoded through the choice of a background cocycle $N$ representing one unit of Chern class, while no explicit lattice model for the curved base is introduced.

The Hopf example therefore shows that the role of the twisted Villain model is not merely to discretise the complete fibred sigma models when this is possible. More generally, it provides a lattice realisation of the fibre-sector topological data relevant for T-duality, even when the full sigma-model structure is not itself implemented on the lattice.

%%%%%%%%%%%%%%%%%%%%%%%%%%%%%%%%%%%%%%%%%%%%%%
%%%%%%%%%%%%   lattice T-duality   %%%%%%%%%%%
%%%%%%%%%%%%%%%%%%%%%%%%%%%%%%%%%%%%%%%%%%%%%%
\section{Lattice half-gauging and topological defect}
\label{sec:defect}

According to  modern understanding, (generalised) symmetries are associated with the existence of topological operators \cite{Gaiotto:2014kfa}. These are defined on some submanifold $\gamma$ of spacetime, and their topological character is reflected in the fact that the theory's correlation functions are invariant under smooth deformations of $\gamma$. Depending on the type of symmetry, topological operators satisfy different fusion rules. Ordinary symmetries satisfy group-like fusion rules. However, there exist topological operators with more general fusion rules, the most well-known example being the Verlinde lines in rational conformal field theories \cite{Verlinde:1988sn}.  In such cases, some topological operators do not admit an inverse and are therefore called \textit{non-invertible} (see e.g. \cite{Shao:2023gho,Schafer-Nameki:2023jdn} for reviews). Dualities are generically encoded through non-invertible symmetries and associated topological defects.

In the previous sections, we constructed the twisted Villain model and established its T-duality at the level of the lattice action. We now turn to the corresponding defect implementation of this duality. Our goal in this section is to construct the lattice T-duality defect by half-gauging and then show that it is topological, namely that the partition function is invariant under deformations of the defect line. T-duality defects on the lattice were previously constructed for the modified Villain model \cite{Choi:2021kmx,Pace:2024oys}, and here we generalise that half-gauging construction to the twisted Villain model. The new ingredient is that the lattice action now contains, in addition to the compact fibre field, cochains encoding the non-trivial fibration and the fibre-horizontal $b$-coupling. These data must therefore be incorporated consistently into the defect construction. We proceed in two steps. First, we implement half-gauging and derive the defect action. Second, we prove that the resulting defect is topological.

%%%%%%%%%%%%%%%%%%%%%%%%%%%%%%%%%%%%%%%%%%%%%%
%%%%%%%%%%%%%%%%%%%%%%%%%%%%%%%%%%%%%%%%%%%%%%
\subsection{Half-gauging the twisted Villain model and defect term}
\label{sec:halfgauging}
The idea of half-gauging \cite{Choi:2021kmx,Choi:2022zal} is to divide spacetime into two regions and gauge a subgroup $\mathbb{Z}_{p}$ of a global symmetry in one of them. In our case, the global symmetry is the isometry symmetry. After gauging, the theories in the two regions will in general be different, and the object residing on  defect locus $\gamma$ will define an interface between them. If, however, a self-duality exists, then for specific values of the free parameters the two theories become equivalent. In that case, the operator on $\gamma$ is a topological defect separating two dual but equivalent descriptions of the same theory.

Let us consider the twisted Villain model \eqref{eq:emV_actioncomplete} on the lattice $\G$. In order to perform half-gauging, let us divide it into two subregions $\Gamma_+$ and $\Gamma_-$, separated by an oriented line $\gamma$, with
\begin{equation}
\label{eq:Sminus_topological}
\partial \Gamma_-=\gamma\ ,\qquad \partial \Gamma_+=-\gamma\ .
\end{equation}
Accordingly, we add a superscript $\pm$ to all cochains to indicate whether they live in $\G_{+}$ or $\G_{-}$. 
The argument presented in this section only require the fibre sector, so let us focus on the fibre-dependent part of the action and split it as 
\be
S_{{\rm fib}} = S_{\G_{-}} + S_{\G_{+}}  \ .
\ee
On $\G_{-}$, we place the twisted Villain model with action
\ba
\label{eq:NLSMactioncupHhalf}
S_{\G_{-}} &=& \frac{R^{2}}{4 \pi } \sum_{\text{links}} \pr{\D \phi^{(-)} - 2 \pi n^{(-)} -  a^{(-)}}^{2} + i \sum_{{\rm plaq}} \pr{\D n^{(-)} - N^{(-)}} \cup    \tilde \phi^{(-)} \nb \\
&+&
 i  \sum_{{\rm plaq}} \pq{\frac{1}{2 \pi} \pr{\D  \phi^{(-)} - 2 \pi  n^{(-)}} \cup v^{(-)} 
 + \phi^{(-)} \cup \tilde N^{(-)}} \ ,
\ea
where the links and the plaquettes in the sums belong to $\G_{-}$. On $\G_{+}$, we place the half-gauged theory obtained by gauging a $\mathbb{Z}_{p}$ subgroup of the isometry symmetry.\footnote{As discussed in section \ref{sec:twistedVillainmodel}, depending on the Chern class of the dual bundle, the isometry symmetry is either $U(1)$ or $\mathbb{Z_{\kappa}}$. In the latter case, when performing half-gauging, we must assume that $p$ is a divisor of $\kappa$.} Applying the same half-gauging procedure as in \cite{Choi:2021kmx} to the twisted Villain model, we obtain on $\Gamma_+$ the action
\ba
\label{eq:NLSMactioncupHhalf2}
S_{\G_{+}} &=& \frac{R^{2}}{4 \pi } \sum_{\text{links}} \pr{\D \phi^{(+)} - 2 \pi n^{(+)}  - \frac{2 \pi}{p} \hat n^{(+)} -   a^{(+)}}^{2} + \frac{2 \pi i}{p} \sum_{{\rm plaq}} \D \hat n^{(+)} \cup m ^{(+)}\nb \\ 
&+&
i  \sum_{{\rm plaq}} \pq{ \frac{1}{2 \pi} \pr{\D  \phi^{(+)} - 2 \pi  n^{(+)} - \frac{2 \pi}{p} \hat n^{(+)}} \cup v^{(+)} - \phi^{(+)} \cup \tilde N^{(+)}} \nb \\
&+& i \sum_{{\rm plaq}} \pr{ \D n^{(+)} -  N^{(+)} + \frac{1}{p} \D \hat n^{(+)} } \cup    \tilde \phi^{(+)} \ ,
\ea
while $S_{\G_{-}}$ remains as in (\ref{eq:NLSMactioncupHhalf}). The partition function of the theory is defined as a product of integrals (for real-valued cochains) and sums (for integer-valued cochains) over the dynamical variables at each lattice site (for 0-cochains) and on each link (for 1-cochains). The variables living on the line $\gamma$ are, so far, doubled, as they appear in both actions. We therefore need to impose matching conditions to eliminate half of them. The basic cochains are not affected by the gauging, so they coincide at $\g$,
\be\label{eq:latticehalfgaungingmatchcond}
a^{(+)}|_{\gamma} = a^{(-)}|_{\gamma} \ , \quad \quad
v^{(+)}|_{\gamma} = v^{(-)}|_{\gamma} \ .
\ee
For the cochains of the fibre sector, we impose
\be
\label{eq:latticehalfgaungingbdycond}
  \phi^{(+)} |_{\g} = \phi^{(-)} |_{\g} \ , \q \q  n^{(+)}_{\g} = n^{(-)}_{\g} \ , \q \q   \tilde\phi^{(+)}|_{\g} = \tilde\phi^{(-)} |_{\g} \  , \quad  \q   \hat n^{(+)} |_{\g} =  0  \ .
\ee 
This action (\ref{eq:NLSMactioncupHhalf2}) is invariant under the following gauge transformations, 
\begin{subequations}
\label{eq:gaugeinvZpgauging}
\ba
\phi^{(+)}&\ra&\phi^{(+)}+ 2 \pi k  + \frac{2\pi}{p} q \ , \\
n^{(+)} &\ra& n^{(+)} + \D k - l \ , \\ 
\widetilde \phi^{(+)}&\ra&\widetilde\phi^{(+)}+2\pi \widetilde k \ , \\
\label{eq:hatngaugetransf}
\hat n^{(+)}&\ra&\hat n^{(+)}+\D q + p l \ , \\
m^{(+)}  &\ra& m^{(+)} - \widetilde k + p \widetilde q\ .
\ea 
\end{subequations}
Here, $k$, $\widetilde k$, $q$ and $\widetilde q$ are integer-valued 0-cochains, whereas $l$ is an integer-valued 1-cochain. 
All these gauge parameter cochains vanish in $\g$. Notice that the last term in the second line of (\ref{eq:NLSMactioncupHhalf2}) is gauge-invariant under the $q$ transformation only if $\widetilde N$ is a multiple of $p$, consistently with our assumptions.

Integrating by parts in the last term of (\ref{eq:NLSMactioncupHhalf2}) and
setting $n^{(+)}$ and $m^{(+)}$ to zero using the invariance under $l$ and $\tilde k$, we obtain
\ba
\label{eq:NLSMactioncupHhalf3}
S_{\G_{+}} &=& \frac{R^{2}}{4 \pi } \sum_{\text{links}} \pr{\D \phi^{(+)}    - \frac{2 \pi}{p} \hat n^{(+)} -   a^{(+)}}^{2}
+ i \sum_{{\rm plaq}} \pr{  \frac{1}{p}  \hat n^{(+)} \cup  \D  \tilde \phi^{(+)} -  N^{(+)} \cup    \tilde \phi^{(+)}}
\nb \\ 
&+&
i  \sum_{{\rm plaq}} \pq{ \frac{1}{2 \pi} \pr{\D  \phi^{(+)}   - \frac{2 \pi}{p} \hat n^{(+)}} \cup v^{(+)} - \phi^{(+)} \cup \tilde N^{(+)}} +  i  \sum_{\g} n^{(-)} \cup \tilde \phi^{(+)} \ ,
\ea
where we also used (\ref{eq:latticehalfgaungingbdycond}). Defining $  \f = p  \phi$ and $ \tilde \f = \tilde \phi /p$ nad relabeling $\hat n \ra n$, we write
\ba
\label{eq:NLSMactioncupHhalf4}
S_{\G_{+}} &=& \frac{R^{2}}{4 \pi p^{2} } \sum_{\text{links}} \pr{\D \f^{(+)}  - 2 \pi    n^{(+)} -   p a^{(+)}}^{2} 
+ i \sum_{{\rm plaq}} \pr{     n^{(+)} \cup  \D  \tilde \f^{(+)} - p N^{(+)} \cup    \tilde \f^{(+)}} \nb \\
&+&
\frac{i }{p} \sum_{{\rm plaq}} \pq{ \frac{1}{2 \pi} \pr{\D  \f^{(+)} - 2 \pi  n^{(+)}} \cup v^{(+)} - \f^{(\pm)} \cup \tilde N^{(\pm)}} + i p \sum_{\g} n^{(-)} \cup \tilde \f^{(+)}\ . 
\ea 

Performing Poisson resummation, and referring to appendix \ref{app:Tdualitycomputations} for the intermediate steps, one obtains the T-dual form
\ba
S_{\G_{+}} 
&=& \frac{p^{2}}{4\pi R^{2} }\sum_{\text{links}} \pr{\D \tilde \f^{(+)}  -  2 \pi \tilde  n^{(+)} - \frac{1}{p}  v^{(+)}}^2 - i  \sum_{{\rm plaq}} \D \f^{(+)} \cup \tilde n^{(+)}  \nb \\  
&-& \frac{i  p}{2 \pi} \sum_{{\rm plaq}} a^{(+)} \cup \pr{\D \tilde \f^{(+)} - \frac{1}{p} v^{(+)} - 2 \pi \tilde n^{(+)}} -   \frac{i}{2 \pi}  \sum_{\g} \pr{ \D \f^{(+)} -  2 \pi p n^{(-)}} \cup \tilde \f^{(+)}  \nb \\
&-& i \sum_{{\rm plaq}} \pr{  p N^{(+)} \cup \tilde \f^{(+)} + \frac{1}{p}  \f^{(+)} \cup \tilde N^{(+)} } \ .
\ea
Rearranging terms and using (\ref{eq:latticehalfgaungingbdycond}), we find
\ba
\label{eq:NLSMactioncupHotherhalf}
S_{\G_{+}} 
&=& \frac{p^{2}}{4\pi R^{2} }\sum_{\text{links}} \pr{\D \widetilde \f^{(+)}  -  2 \pi \widetilde  n^{(+)} - \frac{1}{p}  v^{(+)}}^2 + i  \sum_{{\rm plaq}}  \f^{(+)} \cup \pr{ \D \tilde n^{(+)} - \frac{1}{p} \tilde N^{(+)}}  \nb \\  
&-& i  p  \sum_{{\rm plaq}} \pq{ \frac{1}{2 \pi} a^{(+)} \cup \pr{\D \widetilde \f^{(+)} - \frac{1}{p} v^{(+)} - 2 \pi \widetilde n^{(+)}} +  N^{(+)} \cup \widetilde \f^{(+)} }  \nb \\
&-&  \frac{i p}{2 \pi}  \sum_{\g} \pq{ \pr{ \D \phi^{(-)} -  2 \pi  n^{(-)}} \cup \widetilde \f^{(+)} + 2 \pi \phi^{(-)} \cup \widetilde n^{(+)}} \ .
\ea
This is indeed the twisted Villain model in the dual formulation (\ref{eq:emV_actioncompletedual}), where the last line is an additional term living on the defect locus $\g$. When the non-linear sigma model has target space fibration with Chern number $\l$ and $H$-flux $\kappa$, the cochains $a$ and $N$ are proportional to $\l$, whereas the cochains $v$  and $\widetilde N$ are proportional to $\kappa$. Taking this account, comparing (\ref{eq:NLSMactioncupHhalf}) and (\ref{eq:NLSMactioncupHotherhalf}), one sees that when 
\be
\label{eq:condhalfgauging}
R^{2} = p = \frac{\kappa}{\l} \ , \q \q \q \l, \kappa \neq 0 \ ,
\ee
the theories on $\G_{\pm}$ are self-dual to each other. In the special case $\lambda=\kappa=0$, corresponding to the modified Villain model in which the bundle and flux data vanish, this reduces simply to the condition $R^{2}=p$. 

The conditions (\ref{eq:condhalfgauging}) coincide with the relations found performing half-gauging in the continuum non-linear sigma model \cite{Arias-Tamargo:2025xdd}. Since in the general fibred case T-duality involves topological data as well, it is natural that the topological numbers $\lambda$ and $\kappa$ appear in the condition for half-gauging. Our lattice realisation reproduces precisely these continuum conditions.

Recently, a generalisation of half-gauging for T-duality in the compact boson model, valid also for irrational values of the radius $R$, was proposed both in the continuum \cite{Argurio:2024ewp} and on the lattice \cite{Argurio:2026txf}. In the twisted Villain model, however, the half-gauging condition \eqref{eq:condhalfgauging} ties the radius to the integer topological data $\lambda$ and $\kappa$. This suggests that any analogous extension in the fibred setting would require a different mechanism.

Once (\ref{eq:condhalfgauging}) are satisfied, the  last term in \eqref{eq:NLSMactioncupHotherhalf}, namely
\be
\label{eq:actiondefect}
S_{\g} = - \frac{i p}{2 \pi}  \sum_{\g} \pq{ \pr{ \D \phi^{(-)} -  2 \pi  n^{(-)}} \cup \widetilde \f^{(+)} + 2 \pi \phi^{(-)} \cup \widetilde n^{(+)}} \ ,
\ee
can be interpreted as the insertion in $\g$ of a topological operator in a single non-linear sigma model defined on the whole lattice $\G$. Note that the result (\ref{eq:actiondefect}) matches the form expect from the T-duality defect of  the continuum non-linear sigma models \cite{Niro:2022ctq,Demulder:2022nlz,Arias-Tamargo:2025xdd}.

%%%%%%%%%%%%%%%%%%%%%%%%%%%%%%%%%%%%%%%%%%%%%%
%%%%%%%%%%%%%%%%%%%%%%%%%%%%%%%%%%%%%%%%%%%%%%
\subsection{Partition function invariance under deformation of the defect}

We are left to show that the defect is topological, namely that the partition function is left invariant under a deformation of its locus. The defect is constructed by gauging the $\mathbb Z_p$ symmetry only on the $\Gamma_+$-side of the lattice. Before Poisson resummation, this gives an interface $\g$ of the form
\begin{equation}
	\text{original twisted Villain theory} \;\Big|\; \text{$\mathbb Z_p$-gauged twisted Villain theory}\ .
\end{equation}
We verify the topological character of the interface in the half-gauged formulation. Such a derivation was carried out for the modified Villain model in~\cite{Choi:2022zal}. The relevant steps extend straightforwardly to the twisted Villain model, allowing us to closely follow the same line of reasoning while highlighting the novel features specific to the twisted Villain construction.

We denote by $Z_\gamma$ the full partition function on $\Gamma=\Gamma_-\cup\Gamma_+$, with interface $\gamma$. As in the continuum treatment of duality defect \cite{Arias-Tamargo:2025fhv}, it is convenient to view the partition function on the $+$ side as a functional of the boundary data induced on the interface from the $-$ side.  Writing $\chi^{(\pm)}$ for the collective cochains on $\Gamma_\pm$, this gives
\begin{equation}\label{eq:fullZ_step1}
Z_\gamma = \sum_{\chi^{(-)}} e^{-S_{\Gamma_-}[\chi^{(-)}]} \left( Z_{+,\gamma} \right)\Big|_{\chi^{(+)}|_\gamma=\chi^{(-)}|_\gamma}\ ,
\end{equation}
where the cochains $\chi^{(+)}$ in $\g$ are constrained by the boundary conditions \eqref{eq:latticehalfgaungingmatchcond} and \eqref{eq:latticehalfgaungingbdycond}, and
\begin{equation}\label{eq:Zplus_def}
Z_{+,\gamma} = \sum_{\chi^{(+)}} e^{-S_{\Gamma_+}[\chi^{(+)}]} \ .
\end{equation}
Equivalently, writing $\chi_\gamma$ for the collective interface data, one may express the same partition function as
\begin{equation}\label{eq:fullZ_step2}
Z_\gamma = \sum_{\chi_\gamma} \sum_{\substack{\chi^{(-)}\\ \chi^{(-)}|_\gamma=\chi_\gamma}} e^{-S_{\Gamma_-}[\chi^{(-)}]} \sum_{\substack{\chi^{(+)}\\ \chi^{(+)}|_\gamma=\chi_\gamma}} e^{-S_{\Gamma_+}[\chi^{(+)}]} = \sum_{\chi_\gamma} Z_{-,\gamma}[\chi_\gamma]\,Z_{+,\gamma}[\chi_\gamma] \ .
\end{equation}
The actions $S_{\G_{-}}$ on $\G_-$ and $S_{\G_{+}}$ on $\G_+$ are respectively given by \eqref{eq:NLSMactioncupHhalf} and \eqref{eq:NLSMactioncupHhalf2}. The latter involves a discrete BF theory built from $\hat n^{(+)}$ and $\hat m^{(+)}$ 
\begin{equation}
	S_{\mathrm{BF}} = \frac{2\pi i}{p} \sum_{{\rm plaq}\subset\Gamma^{(+)}} \hat m^{(+)}\cup\Delta \hat n^{(+)} \ ,
\end{equation}
where the cochain $\hat n^{(+)}$ transforms according to (\ref{eq:hatngaugetransf}). The corresponding BF partition function is
\begin{align}\label{eq:ZBF}
	Z_{\mathrm{BF},\gamma} = \frac{1}{p^{\#(2)}} \frac{1}{p^{\#(0)}} \sum_{\widehat n^{(+)},\widehat m^{(+)}} \exp\left[ \frac{2\pi i}{p} \sum_{{\rm plaq}\subset\Gamma_+(\gamma)} \widehat m^{(+)}\cup\Delta \widehat n^{(+)} \right]\ ,
\end{align}
where $\#(r)$ is the number of $r$-cells in the gauged region of the lattice. The factor $p^{-\#(2)}$ divides by the volume of the plaquette multiplier variables $\widehat m^{(+)}$, while $p^{-\#(0)}$ divides by the volume of the $\mathbb Z_p$ gauge transformations generated by the $0$-cochain $q$ in \eqref{eq:hatngaugetransf}.

%%%%%%%%%%%%%%%%%%%%%%%%%%%%%%%%%%%
%-%-%-%-%-%-%-%-%-%-%-%-%-%-%-%-%-%
%%%%%%%%%%%%%%%%%%%%%%%%%%%%%%%%%%%
\begin{figure}[t]
	\centering
	\begin{tikzpicture}[scale=0.85]
		
		% grid
		\draw[step=1cm,gray!45,very thin] (-1.9,-1.9) grid (2.9,2.9);
		
		% shade Gamma_+
		\fill[gray!10] (0,-1.9) rectangle (2.9,2.9);
		
		% interface gamma
		\draw[red, thick] (0,-2) -- (0,2);
		\draw[red, thick] (0,2) -- (0,3) node[midway,right] {$\gamma$};
		
		% redraw grid on top of shading
		\draw[step=1cm,gray!45,very thin] (-1.9,-1.9) grid (2.9,2.9);
		
		% open-circle vertices
		\foreach \x in {-1,0,1,2}
		\foreach \y in {-1,0,1,2}
		\draw[fill=white,draw=black] (\x,\y) circle (2.1pt);
		
		% labels
		\draw[red,thick] (0,0.1) -- (0,0.9) node[pos=0.55, left] {$\hat n_0$};
		\draw[very thin,gray!45] (1,0) -- (1,1) node[pos=0.55, left] {\color{black}$\hat m_0$};
		\draw[black,thin](0,0) node[pos=0.55, left] {\color{black}$x$};
		\node[left] at (-1.2,2.5) {$\Gamma_-$};
		\node[right] at (2.2,2.5) {$\Gamma_+$};
	\end{tikzpicture}
	\hspace{1.5cm}
	\begin{tikzpicture}[scale=0.85]
		% grid
		\draw[step=1cm,gray!45,very thin] (-1.9,-1.9) grid (2.9,2.9);
		
		% shade Gamma_+'
		\fill[gray!10] (0,-1.9) rectangle (2.9,2.9);
		\fill[white] (0,0) rectangle (1,1); % moved plaquette now removed from Gamma_+'
		
		% redraw grid on top
		\draw[step=1cm,gray!45,very thin] (-1.9,-1.9) grid (2.9,2.9);
		
		% deformed interface gamma'
		\draw[red, thick] (0,-2) -- (0,0);
		\draw[red, thick] (0,0) -- (1,0);
		\draw[red, thick] (1,0) -- (1,1);
		\draw[red, thick] (1,1) -- (0,1);
		\draw[red, thick] (0,1) -- (0,2);
		\draw[red, thick] (0,2) -- (0,3) node[midway,right] {$\gamma'$};
		
		% open-circle vertices
		\foreach \x in {-1,0,1,2}
		\foreach \y in {-1,0,1,2}
		\draw[fill=white,draw=black] (\x,\y) circle (2.1pt);

		% labels on links / gauge parameters
		\draw[red] (0.1,0) -- (0.9,0) node[midway, below] {$\hat n_3$};
		\draw[red] (0.1,1) -- (0.9,1) node[midway, above] {$\hat n_1$};
		\draw[red] (1,0.1) -- (1,0.9) node[midway, right] {$\hat n_2$};
			\draw[black,thin](0,0) node[pos=0.55, left] {\color{black}$x$};
		\node[above right] at (1,1) {$q_1$};
		\node[below right] at (1,0) {$q_3$};
		\draw[very thin,gray!45] (1,0) -- (1,1) node[pos=0.5,left] {\color{black}$p_0$};
		\node[left] at (-1.2,2.5) {$\Gamma_-'$};
		\node[right] at (2.2,2.5) {$\Gamma_+'$};
	\end{tikzpicture}
	\caption{Local deformation of the interface. On the left, $\gamma$ separates $\Gamma_-$ from $\Gamma_+$. On the right, $\gamma'$ is obtained by moving the interface across the central plaquette. The shaded region indicates the gauged side. The boundary links of the plaquette $p_0$ are labeled clockwise as $0$, $1$, $2$ and $3$. }
	\label{fig:plaquette_collars}
\end{figure}
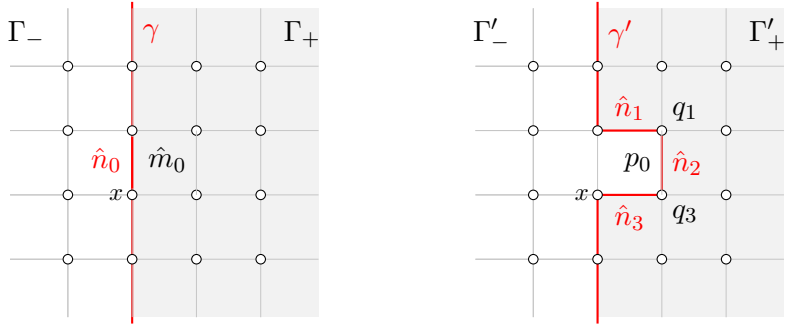
%%%%%%%%%%%%%%%%%%%%%%%%%%%%%%%%%%%
%-%-%-%-%-%-%-%-%-%-%-%-%-%-%-%-%-%
%%%%%%%%%%%%%%%%%%%%%%%%%%%%%%%%%%%

To prove that the defect is topological, it is enough to show that the partition function is invariant under an elementary local deformation of the defect line, namely $Z_{\gamma}=Z_{\gamma'}$. We therefore consider the move in which a single plaquette $p_0$ is transferred from $\Gamma_+$ to $\Gamma_-$. Since an arbitrary deformation of the defect can be decomposed into a sequence of such elementary plaquette moves, this local check proves the general statement.

On the half-gauged side, this move is implemented by integrating out the BF multiplier $\hat m_0$ on the plaquette $p_0$, and then gauge-fixing so that the boundary condition $\hat n^{(+)}=0$ is imposed on the new interface $\gamma'$, namely on the links labelled $1$, $2$, and $3$ in figure \ref{fig:plaquette_collars}. Integrating out $\hat m_0$ produces a factor of $p$ in $Z_{\gamma,+}$ and imposes the local flatness condition
\begin{equation}
\label{eq:flatness condition}
\hat n_3^{(+)}+\hat n_2^{(+)}-\hat n_1^{(+)}=0 \quad \bmod p \ ,
\end{equation}
where we used the boundary condition \eqref{eq:latticehalfgaungingbdycond} on the original interface $\gamma$, for which $\hat n_0^{(+)}=0$. The relation \eqref{eq:flatness condition} then allows  to gauge-fix
\begin{equation}
	\hat n_1^{(+)}=\hat n_2^{(+)}=\hat n_3^{(+)}=0 
\end{equation}
while using only two $\mathbb Z_p$ gauge redundancies \cite{Choi:2022zal}. This produces an additional factor of $p^2$ in the partition function.

As a result, under these manipulations, the BF part of the partition function $Z_{+,\g}$ is modified according to 
\begin{align}\label{eq:explicit_BF_move_gamma_gammaprime}
		Z_{\rm BF,\gamma}&=\frac{1}{p^{\#(2)}}\frac{1}{p^{\#(0)}}\sum_{\hat n,\hat m}\exp\left(\frac{2\pi i}{p}\sum_{\rm plaq}\hat m^{(+)}\cup\Delta \hat n^{(+)} \right)\notag \\
		&= \frac{1}{p^{\#(2)}}\frac{1}{p^{\#(0)}}p\,p^2\sum_{\hat n,\hat m\setminus \hat m_0} \exp\left( \frac{2\pi i}{p}\sum_{\rm plaq}\hat m^{(+)}\cup\Delta \hat n^{(+)} \right)\times \delta_{\hat n_1^{(+)} = 0}\, \delta_{\hat n_2^{(+)} = 0}\, \delta_{\hat n_3^{(+)} =  0}\notag \\
		&=Z_{\rm{BF},\g'}\,\delta_{\hat n_1^{(+)} = 0}\, \delta_{\hat n_2^{(+)} = 0}\, \delta_{\hat n_3^{(+)} =  0}\ .
\end{align}
Here, the Kronecker deltas encode the new boundary conditions on $\gamma'$ after gauge-fixing. The factor of $p$ arising from the sum over $\widehat m_0$ and the factor of $p^2$ associated with gauge-fixing the two newly exposed $\widehat n$-links are exactly compensated by the change in the BF normalization, since after the move the gauge region contains one fewer plaquette and two fewer gauge parameters.

At fixed interface data, the contribution from the gauged side may therefore be rewritten as
\begin{equation}
    Z_\g=Z_{\g,-}\sum_{\hat n^{(+)}} Z^{\rm rest}_{\G_+}[\hat n^{(+)}]\,\delta_{\hat n_1^{(+)} = 0}\, \delta_{\hat n_2^{(+)} = 0}\, \delta_{\hat n_3^{(+)} =  0}\,Z_{\g',\rm BF}[\hat n^{(+)}]\ ,
\end{equation}
where $Z^{\rm rest}_{\G_+}[\hat n^{(+)}]$ denotes the part of the $\Gamma_+$-partition function that is independent of the BF-multiplier $\widehat m^{(+)}$. Since the cochain $\hat n^{(+)}$ vanishes on the boundary links of $p_0$, this contribution factorises into a term on $\Gamma_+'$ and a local plaquette weight on $p_0$, now defined as a plaquette in $\G_-'$,
\ba 
Z^{\rm rest}_{\G_+}[\hat n^{(+)}]\,\delta_{\hat n_1^{(+)} = 0}\, \delta_{\hat n_2^{(+)}=0}\delta_{\hat n_3^{(+)} =  0}&=&Z^{\rm rest}_{\G_+'}[\hat n^{(+)}]\,Z^{\rm rest}_{\G_+}[\hat n^{(+)}]|_{p_0\subset\G_+}\nb\\
&=& Z^{\rm rest}_{\G_+'}[\hat n^{(+)}]\,Z^{\rm rest}_{\G_-'}|_{p_0\subset\G_-'}\ .
\ea
Hence, we may relabel the other fields $\chi^{(+)}$, defined on the plaquette $p_0$ or on its boundary boundary links, by $\chi^{(-)}$ while imposing the matching conditions 
\be \chi^{(-)}|_{\g'}=\chi^{(+)}|_{\g'} \ .\ee

Let us consider, in particular, the contribution of the $b$-term in $Z^{\rm rest}_{\G_+}[\hat n^{(+)}]$. Using the new boundary constraint, $\hat n_1^{(+)}= \hat n_2^{(+)} = \hat n_3^{(+)}=0$, the contribution of the plaquette $p_0$ in the partition function reads
\begin{align}
    (Z^{\rm{rest}}_{\G_+,b})|_{p_0}&=\left.\exp{\pr{-i\pq{ \frac{1}{2 \pi} \pr{\D  \phi^{(+)} - 2 \pi  n^{(+)} - \frac{2 \pi}{p} \hat n^{(+)}} \cup v^{(+)} - \phi^{(+)} \cup \tilde N^{(+)}}}}\right|_{p_0}\nb\\
    &=\exp{\pr{-\frac{i}{2\pi}\pq{\pr{\D_3\phi-2\pi n_3}^{(+)}v_2^{(+)}-\pr{\D_0\phi-2\pi n_0}^{(+)}v_1^{(+)}-\phi_x^{(+)}\tilde N^{(+)}_{x,12}}}} \nb \ ,
\end{align}
where $\Delta_i\phi$ denotes the lattice difference of $\phi$ along link $i$ and $x$ is the vertex on the bottom left of the paquette $p_0$. This has exactly the same form as the $b$-term in $\G_-$ \eqref{eq:NLSMactioncupHhalf} up to relabelling $\D\phi^{(+)}\ra \D\phi^{(-)}$, $n^{(+)}\ra n^{(-)}$, $v^{(+)}\ra v^{(-)}$, $\widetilde{N}^{(+)}|_{p_0}\ra  \widetilde{N}^{(-)}|_{p_0}$ and imposing the matching conditions
\ba
\phi_{x}^{(+)}=\phi_{x}^{(-)}\ .\q \q  n^{(+)}=n^{(-)}\ .\q \q  v^{(+)}= v^{(-)}
\ea on the new boundary links $1$, $2$, $3$ (see figure \ref{fig:plaquette_collars}). 
Under this identification, the local $b$-term coincides with the $b$-term of the original twisted Villain action on the same plaquette, now regarded as belonging to $\Gamma'_-$
\begin{equation*}
    (Z^{\rm{rest}}_{\G_+,\, b})|_{p_0}=\exp\left[-i\left(\frac{1}{2\pi}\bigl(\Delta\phi^{(-)}-2\pi n^{(-)}\bigr)\cup v^{(-)}-\phi^{(-)}\cup \widetilde N^{(-)}\right)\Big|_{p_0}\right]=(Z^{\rm{rest}}_{\Gamma_-',\, b})|_{p_0}\ .
\end{equation*}
The same local reassignment can be checked for the remaining plaquette contributions, in particular for the Lagrange-multiplier term enforcing the twisted vorticity condition. Hence, we conclude that the local plaquette weight is unchanged when $p_0$ is displaced from $\Gamma_+$ to $\Gamma_-$, thus
\begin{equation}
    Z_\gamma= Z_{\gamma'}\ 
\end{equation} 
and the interface is topological. This demonstrates that the lattice T-duality with flux-bundle exchange constructed in section \ref{sec:twistedVillain} can be realised through a topological interface.

%%%%%%%%%%%%%%%%%%%%%%%%%%%%%%%%%%%%%%%%%%%%%%
%%%%%%%%%%%%      Conclusion      %%%%%%%%%%%%
%%%%%%%%%%%%%%%%%%%%%%%%%%%%%%%%%%%%%%%%%%%%%%
\section{Conclusion} \label{sec:conclusion}

We showed that a genuinely stringy form of T-duality is realised exactly on the lattice. Starting from the modified Villain model of the compact boson, we introduced a twisted version adapted to circle fibrations with background flux data, and showed that it captures the characteristic bundle-flux exchange of T-duality together with its associated topological defect.

Concretely, we defined the twisted Villain model by coupling the lattice fibre field to cochains encoding the bundle connection and the fibre-horizontal component of the B-field, together with a twisted vorticity constraint fixed by the Chern class data. We showed how this construction realises several fibred backgrounds, including toroidal, nilfold and the Hopf-fibration examples at the level of their fiber sector. We then implemented lattice T-duality by half-gauging, recovered the expected exchange of bundle and flux data, and derived the corresponding defect action. Finally, we established that it defines a topological defect, and hence a generalised symmetry, directly on the lattice. 

A crucial lesson in curved backgrounds with non-trivial circle fibrations, the exact lattice realisation of T-duality is no longer organised by the ordinary momentum-winding exchange of the free boson. Instead, it is controlled by the bundle curvature, the fibre-horizontal $B$-field data. Correspondingly, the fibred theory does not require an exact global winding symmetry: beyond the free boson, the relevant input is the fiber isometry together with the associated bundle-flux exchange. 

More broadly, our results show that the exact lattice realisations of stringy T-duality need not require a complete discretisation of the underlying non-linear sigma-model. In the continuum, the relevant topological data is intrinsic to the target-space geometry; bundle curvature and flux are encoded geometrically already off-shell. In the lattice model, by contrast, there is no underlying smooth geometry from which this structure is inherited. Instead, the corresponding bundle and flux data are sourced through the vorticity arising from the Villain cochains. It is precisely in this sense that the twisted Villain model is the key to provide a lattice avatar for the bundle-flux exchange of fibred T-duality.

\bigskip

\noindent
Several natural directions remain open. 
\vspace{-6pt}

\paragraph{Lattice realisation with curved base-manifolds.}
A natural direction is to extend the present construction so as to treat backgrounds for which the base is itself genuinely curved, such as the $S^2$ base of the $SU(2)$ WZW model. In such cases, the present Abelian cochain formalism is unlikely to be sufficient: one expects the relevant lattice data to involve also non-Abelian cochains or similar generalised discrete structure capable of encoding curved bundle data together with the corresponding flux sector. From this viewpoint, the obstacle is not simply that the computation becomes more involved, but that the appropriate lattice language should itself be enlarged. Developing such a framework would open the way to lattice realisations of a much broader class of curved sigma-model backgrounds. Related, it would also be interesting to understand whether a future extension of the present framework, incorporating a fuller lattice sigma-model dynamics, could be related to discrete descriptions of classical string motion such as segmented strings in  $\mathrm{AdS}_3$, see e.g. \cite{Callebaut:2015fsa,Gubser:2016wno,Gubser:2016zyw,Vegh:2016hwq,Vegh:2021jhl}.

\paragraph{Lattice correspondence space and topological T-duality.}
Even within the present Abelian fibre-sector framework, it is natural to ask how far the construction can be pushed toward a genuine lattice version of topological T-duality \cite{Bouwknegt:2003wp,Bouwknegt:2003vb}. In this work, we isolated and realised exactly the fibre-sector data responsible for the bundle-flux exchange, but this does not yet provide the full correspondence-space formalism. In particular, we only control the fibrewise part of the $B$-field sector, whereas the basic flux components and their compatibility with the bundle data remain to be incorporated. More generally, such an extension would have to account for the fact that the lattice is governed by cochain calculus rather than differential calculus, so the familiar continuum descent relations are expected to acquire genuinely lattice-specific corrections. Clarifying whether the full topological-T-duality apparatus admits a natural cochain-theoretic reformulation therefore remains an important open problem.

\paragraph{Anomalies.}
An important question regarding the global symmetries of any model is whether they are characterised by 't Hooft anomalies \cite{tHooft:1979rat}. The compact boson model exhibits a mixed anomaly between the momentum and winding symmetries $U(1)$, and the modified Villain model was shown to reproduce them \cite{Gorantla:2021svj}. 
The non-linear sigma models studied in this work display an interesting pattern of 't Hooft anomalies \cite{Arias-Tamargo:2025fhv}. It would therefore be interesting to determine whether the twisted Villain model reproduces not only the symmetry pattern of these fibred theories, but also their associated 't Hooft anomalies in a precise lattice cochain formulation.

\paragraph{Finite-depth circuit realisation?}
It would also be interesting to understand whetherc the present defect construction admits a natural Hamiltonian interpretation in terms of local or finite-depth circuit transformations. Recent work on exact lattice T-duality in quantum spin chains has shown that non-invertible symmetry provides a useful reformulation for such questions \cite{Pace:2024oys}. In the fibred models studied here, however, the accompanying bundle-flux exchange suggests a subtler notion of equivalence than in the compact-boson case. A Hamiltonian understanding of the defect could help clarify whether these dual descriptions are related by shallow local transformations, or whether the duality is intrinsically mediated by more non-local topological data.

\bigskip
%%%%%%%%%%%%%%%%%%%%%%%%%%%%%%%%%%%%%%%%%%%%%%
\subsubsection*{Acknowledgments}
We thank Riccardo Argurio, Rathindra Nath Das and Christian Northe for useful discussions.
C.B., A.C.~and J.E.~acknowledge the support of the German Research Foundation (DFG) through the
Collaborative Research Center ToCoTronics, Project-ID 258499086 — SFB 1170, as well
as Germany’s Excellence Strategy through the Würzburg-Dresden Cluster of Excellence on
Complexity and Topology in Quantum Matter - ctd.qmat (EXC 2147, Project-ID 390858490).

\newpage
%%%%%%%%%%%%%%%%%%%%%%%%%%%%%%%%%%%%%%%%%%%%%%
%%%%%%%%%%%%        Appendix      %%%%%%%%%%%%
%%%%%%%%%%%%%%%%%%%%%%%%%%%%%%%%%%%%%%%%%%%%%%
\appendix
\addtocontents{toc}{\protect\setcounter{tocdepth}{1}}

%%%%%%%%%%%%%%%%%%%%%%%%%%%%%%%%%%%%%%%%%%%%%%
%%%%%%%%%%%%        cochain       %%%%%%%%%%%%
%%%%%%%%%%%%%%%%%%%%%%%%%%%%%%%%%%%%%%%%%%%%%%
\section{Cochains technology}\label{app:cochains}
In this appendix, we review some basic facts about cochains and the manipulations involving them. This also allows us to establish the notation used throughout the paper.

The cochain formalism provides a natural discrete framework for describing lattice matter and gauge fields. We work with an Euclidean two dimensional square lattice $\G$. On such a lattice, we find the following subspaces, usually called \textit{cells}:
\begin{align}
\text{0-cells (sites)} \ , \q \q 
\text{1-cells (links)} \ , \q \q 
\text{2-cells (plaquettes)} \ . 
%\text{3-cells (cubes)},\ \ldots
\end{align}
Cochains are the discrete analogue of the differential forms. A $p$-cochain $c$ assigns a number to each oriented $p$-cell. 
We therefore have $0$-cochains defined on sites, $1$-cochains defined on links and $2$-cochains defined on plaquettes. When working in components, we write
\begin{align}
\text{0-cochain:} \q \phi_{x}   \ ,  \q \q
\text{1-cochain:} \q n_{x,\mu}  \ , \q \q 
\text{2-cochain:} \q b_{x,\mu \nu} = - b_{x,\nu \mu} \ . 
\end{align}
Here, $x$ denotes the lattice sites, while the indices $\mu$ and $\nu$ take values $1$ or $2$. Thus, $n_{x,\mu}$ is a number associated with the link emanating from the site $x$ in the $\mu$ direction. It is also convenient to introduce the two unit vectors $\hat 1$ and $\hat 2$, so that $x+\hat 1$ denotes the site obtained by moving one lattice spacing along direction $1$, i.e.
\begin{equation}
\raisebox{-0.5\height}{%
   \includegraphics[scale=0.105]{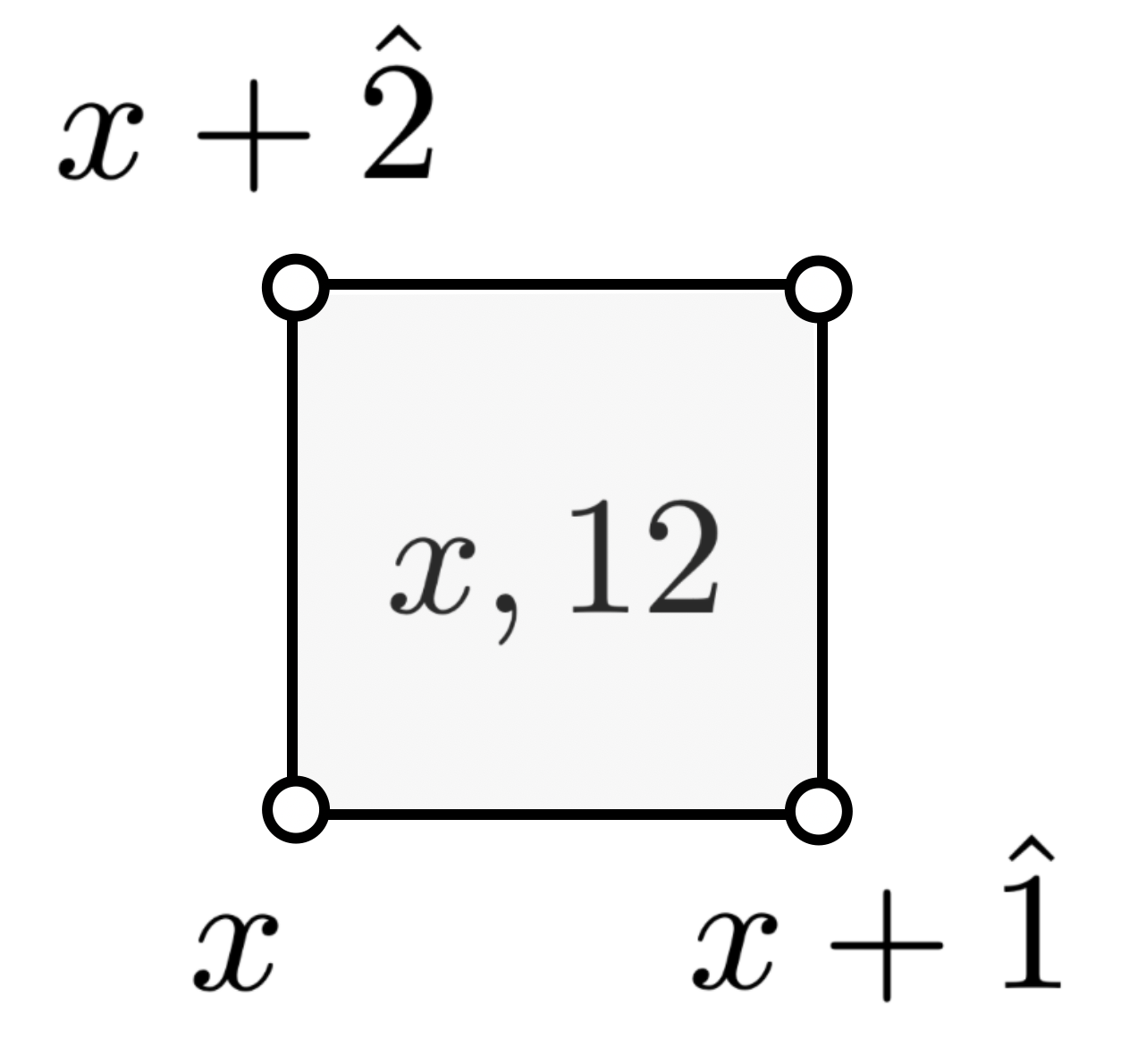}}
\end{equation}
We denote the space of $p$-cochains with values in an abelian group $G$ by $C^p(\G,G)$.

We now describe two important operations involving cochains. 

\paragraph{Coboundary operator.}
The lattice analogue of the exterior derivative is the coboundary operator
\begin{align}
\Delta : \q C^p(\G,G)\to C^{p+1}(\G,G)\ .
\end{align}
For a $0$-cochain $\phi$,
\begin{align}
(\Delta\phi)_{x,\mu}=\phi_{x+\hat\mu}-\phi_x \ .% \equiv \Delta_\mu\phi_x\ .
\end{align}
For a $1$-cochain $n$, its coboundary on an oriented plaquette $(\mu,\nu)$ is the
discrete curl
\begin{align}
(\Delta n)_{x,\mu\nu}=n_{x,\mu}+n_{x+\hat\mu,\nu}-n_{x+\hat\nu,\mu}-n_{x,\nu}\ .\qquad \mu<\nu\ .
\end{align}
As in the continuum,
\begin{align}
\Delta^2=0\ .
\end{align}

\paragraph{Cup product.}
The cubical cup product
\begin{align}
\cup : C^p(\G,G)\times C^q(\G,G')\to C^{p+q}(\G,G\otimes G')
\end{align}
is the lattice analogue of the wedge product. In the notation above, the basic formulas we
need are (see appendix A of \cite{Jacobson:2023cmr})
\begin{subequations}
\begin{align}
(\alpha^{(0)}\cup \beta^{(1)})_{x,1}&=\alpha_x\,\beta_{x,1}\ ,\\
(\beta^{(1)}\cup \alpha^{(0)})_{x,1}&=\beta_{x,1}\,\alpha_{x+\hat 1}\ ,\\
(\alpha^{(1)}\cup \beta^{(1)})_{x,12}&=\alpha_{x,1}\,\beta_{x+\hat 1,2}-\alpha_{x,2}\,\beta_{x+\hat 2,1}\ .\label{eq:cup_product_11_app}\\
(\alpha^{(0)}\cup \beta^{(2)})_{x,12}&=\alpha_x\,\beta_{x,12}\ ,\\
(\beta^{(2)}\cup \alpha^{(0)})_{x,12}&=\beta_{x,12}\,\alpha_{x+\hat 1+\hat 2}\ .
\end{align}
\end{subequations}
For two $1$-cochains, the antisymmetry appropriate to the continuum wedge product is already built into \eqref{eq:cup_product_11_app}; one should therefore not antisymmetrise it again. For example

\begin{align}
(\alpha\cup\beta)_{x,12}&=\alpha_{x,1}\,\beta_{x+\hat 1,2}- \alpha_{x,2}\,\beta_{x+\hat 2,1} \\[1em]
\raisebox{-10pt}{\begin{tikzpicture}[baseline=-0.2ex,scale=0.7]
  % plaquette
  \draw[thick, fill=lightgray!30!white] (0,0) rectangle (1.4,1.4);
  % base point
  \fill (1.4,0) circle (1.3pt);
  \node[below right] at (-0.5,0) {$x$};
  % directions
  \draw[->,thick] (1.4,0.08) -- (1.4,1.4);
  \node[right] at (1.35,0.84) {$\hat 2$};
  \draw[->,thick] (0.1,-0.) -- (1.40,-0.);
  \node[below] at (0.9,0.05) {$\hat 1$};
\end{tikzpicture}}\,&=\;\;
\raisebox{-10pt}{\begin{tikzpicture}[baseline=-0.2ex,scale=0.7]
  % square
  \draw[thick] (0,0) rectangle (1.4,1.4);
  \fill (1.3,0) circle (1.2pt);
  \node[below right] at (-0.5,0) {$x$};
  % alpha on right edge
  \draw[->,thick,draw = purple] (0,0) -- (1.40,0);
  \node[right] at (0.48,-0.35) {$\alpha$};
  % beta on top edge
  \draw[->,thick,draw = purple] (1.4,0) -- (1.4,1.4);
  \node[left] at (2.1,0.8) {$\beta$};
\end{tikzpicture}}
\;-\;
\raisebox{-10pt}{\begin{tikzpicture}[baseline=-0.2ex,scale=0.7]
  % square
  \draw[thick] (0,0) rectangle (1.4,1.4);
  %\fill (1.3,0) circle (1.2pt);
   \node[below right] at (-0.5,0) {$x$};
  % alpha on left edge
  \draw[->,thick,draw = purple] (0.0,0) -- (0.0,1.4);
  \node[left] at (1.3,1.75) {$\beta$};
  % beta on bottom edge
  \draw[->,thick,draw = purple] (0,1.4) -- (1.4,1.4);
  \node[left] at (0.0,0.6) {$\alpha$};
\end{tikzpicture}}
\end{align}

The cup product obeys the Leibniz rule. Considering $\a \in C^{p}(\G,G)$ and $\b \in C^{q}(\G,G)$, this reads
\begin{align}\label{eq:Leibniz_app}
\Delta(\alpha\cup\beta)=\Delta\alpha\cup\beta+(-1)^{p}\alpha\cup\Delta\beta\ .
\end{align}
On a periodic lattice this immediately implies the discrete summation-by-parts identity
\begin{align}\label{eq:sbp_app}
\sum \Delta\alpha\cup\beta=(-1)^{p+1}\sum \alpha\cup\Delta\beta \ ,
\end{align}
where the sum is over any $(p+q+1)$-cycle.

Unlike the continuum wedge product, the cup product is not graded-commutative, 
\begin{align}\label{eq:graded_comm_failure_app}
\alpha\cup\beta-(-1)^{pq}\beta\cup\alpha=(-1)^{p+q+1}\Bigl[\Delta(\alpha\cup_1\beta)-\Delta\alpha\cup_1\beta-(-1)^p\alpha\cup_1\Delta\beta\Bigr] \ , 
\end{align}
for $\alpha\in C^p$, $\beta\in C^q$. On the right-hand side, the first higher-cup product appears. In general, the higher-cup product is an application (see e.g. \cite{Chen:2021ppt} for a combinatorial definition) 
\begin{align}
\cup_i : C^p\times C^q \to C^{p+q-i}\ ,
\end{align}
and vanishes unless $i\leq p,q$. Both the ordinary and the higher-cup products are neither graded-commutative nor associative. However, both properties are retrieve in cohomology.

%%%%%%%%%%%%%%%%%%%%%%%%%%%%%%%%%%%%%%%%%%%%%%
%%%%%%%%%%%%%%%%%%%%%%%%%%%%%%%%%%%%%%%%%%%%%%
\section{Lattice T-duality computations} 
\label{app:Tdualitycomputations}
In this appendix, we detail on the derivation of T-duality for the twisted Villain model. We consider the general setting in which the lattice has a boundary $\g$. This is the case of section \ref{sec:halfgauging}, where half-gauging is discussed. In section \ref{sec:twistedVillain}, we perform T-duality in cases where the lattice has no boundary. To retrieve the formulas there, it is sufficient to send to zero the terms depending on $\g$ in the derivation below. Moreover, setting $a=v=0$ and $N=\widetilde N=0$, one retrieves the derivation of T-duality for the modified Villain model.

For convenience, let us rewrite here the action of the twisted Villain model, 
\ba
\label{eq:twistedmodelactionApp}
    S &=& S_{{\rm bas}} + \frac{R^2}{4\pi}\sum_{\rm{links}} \pr{\D \phi - a - 2 \pi n}^2 + i \sum_{\rm{plaq}} \pr{\D n - N} \cup  \widetilde\phi \nb \\
&+& i \sum_{\text{plaq}} \pq{\frac{1}{2 \pi} \pr{ \D \phi - 2 \pi n  } \cup  v  - \phi \cup \tilde N}  \ .
\ea
On the lattice, T-duality is obtained through the Poisson resummation formula \cite{Sulejmanpasic:2019ytl, Gorantla:2021svj}
\be
\label{eq:Poissonformula}
\sum_{n} \exp \pq{ - \frac{R^{2}}{4 \pi} \pr{\th - 2 \pi n}^{2} - i n \tilde \th  } 
= \frac{1}{R} \sum_{\tilde n} \exp \pq{ - \frac{1}{4 \pi R^{2}} \pr{\tilde \th - 2 \pi  \tilde n}^{2} - \frac{i}{2 \pi} \theta \pr{ \tilde \th  - 2 \pi \tilde n} } \ .
\ee
Integrating by parts, the action (\ref{eq:twistedmodelactionApp}) reads
\ba
\label{eq:NLSMactioncupH2}
S &=& S_{{\rm bas}}  + \frac{R^{2}}{4 \pi } \sum_{\text{links}} \pr{\D \phi - 2 \pi n - a}^{2} + i \sum_{{\rm plaq}}  \pr{  n   \cup \D  \tilde \phi  - N \cup  \tilde \phi}  \nb \\
&+& i \sum_{\text{plaq}} \pq{\frac{1}{2 \pi} \pr{\D  \phi - 2 \pi  n} \cup v  - \phi \cup \tilde N} + i \sum_{\g} n_{\g} \cup \widetilde \phi \ .
\ea
On $\g$, we give boundary conditions, so that $n_{\g} \equiv n|_{\g}$ is fixed and not involved in the Poisson resummation formula. For brevity, let us call $S_{\text{sp}} = S_{{\rm bas}} + i \sum_{\g} n_{\g} \cup \widetilde \phi$. Using the notation described in appendix \ref{app:cochains} ,
the partition function is\footnote{We leave understood the integrations over the base cochains $y^{\m}$ and the sums over the base cochains $n^{(y^{\m})}$, as they are not involved in the following T-duality steps.}
\ba
Z 
&=&  \prod_{x} \int d \phi_{x}  d \tilde \phi_{x} \sum_{n_{x,1}} \sum_{n_{x,2}} e^{-S} \nb \\
&=&  \prod_{x} \int d \phi_{x}  d \tilde \phi_{x} e^{- S_{\text{sp}} + i \phi_{x} \tilde N_{x+\hat 1 + \hat 2, 12} + i N_{x,12} \tilde \phi_{x+\hat 1 + \hat 2}} \times \nb \\
&\times&
\sum_{n_{x,1}} e^{-\frac{R^{2}}{4 \pi }   \pr{\D \phi_{x,1} - 2 \pi n_{x,1} - a_{x,1}}^{2}  - i n_{x,1}\,\D \tilde \phi_{x+\hat 1,2} - \frac{i}{2 \pi} \pr{\D \phi_{x,1} - 2 \pi n_{x,1}} v_{x+\hat 1,2}} \times  \nb \\
&\times& \sum_{n_{x,2}} e^{-\frac{R^{2}}{4 \pi }   \pr{\D \phi_{x,2} - 2 \pi n_{x,2} - a_{x,2}}^{2}  + i n_{x,2}\,\D \tilde \phi_{x+\hat 2,1} + \frac{i}{2 \pi} \pr{\D \phi_{x,2} - 2 \pi n_{x,2}} v_{x+\hat 2,1}} \nb \\
&=&  \prod_{x} \int d \phi_{x}  d \tilde \phi_{x}  e^{- S_{\text{sp}} + i \phi_{x} \tilde N_{x+\hat 1 + \hat 2, 12} + i N_{x,12} \tilde \phi_{x+\hat 1 + \hat 2} -\frac{i}{2 \pi} \pr{\D \phi_{x,1} v_{x+\hat 1,2} - \D \phi_{x,2} v_{x+\hat 2,1}}} \times \nb \\ 
&\times &\sum_{n_{x,1}} e^{-\frac{R^{2}}{4 \pi }   \pr{\D \phi_{x,1} - 2 \pi n_{x,1} - a_{x,1}}^{2}  - i n_{x,1}\pr{ \D \tilde \phi_{x+\hat 1,2} - v_{x+\hat 1,2}} } \times  \nb \\
&\times& \sum_{n_{x,2}} e^{-\frac{R^{2}}{4 \pi }   \pr{\D \phi_{x,2} - 2 \pi n_{x,2} - a_{x,2}}^{2}  + i n_{x,2} \pr{ \D \tilde \phi_{x+\hat 2,1} - v_{x+\hat 2,1}} }  \ .
\ea
We now apply the Poisson formula (\ref{eq:Poissonformula}) both for $n_{x,1}$ and $n_{x,2}$.
\ba
Z 
&=& \frac{1}{R}  \prod_{x} \int d \phi_{x}  d \tilde \phi_{x}  e^{- S_{\text{sp}} + i \phi_{x} \tilde N_{x+\hat 1 + \hat 2, 12} + i N_{x,12} \tilde \phi_{x+\hat 1 + \hat 2} -\frac{i}{2 \pi} \pr{\D \phi_{x,1} v_{x+\hat 1,2} - \D \phi_{x,2} v_{x+\hat 2,1}}}  \nb \\
&\times & \sum_{n_{x+\hat 1,2}} e^{-\frac{1}{4 \pi R^{2} }   \pr{\D \tilde \phi_{x+\hat 1,2} -v_{x+\hat 1,2} - 2 \pi  \tilde n_{x+\hat 1,2}}^{2}  - \frac{i}{2 \pi} \pr{ \D \phi_{x,1} - a_{x,1}}\,\pr{ \D \tilde \phi_{x+\hat 1,2}- v_{x+\hat 1,2} - 2 \pi \tilde n _{x+\hat 1,2}}} \times  \nb \\
&\times & \sum_{n_{x+\hat 2,1}} e^{-\frac{1}{4 \pi R^{2} }   \pr{\D \tilde \phi_{x+\hat 2,1} - v_{x+\hat 2,1} - 2 \pi  \tilde n_{x+\hat 2,1}}^{2}  + \frac{i}{2 \pi} \pr{ \D \phi_{x,2} - a_{x,2}}\,\pr{ \D \tilde \phi_{x+\hat 2,1}-v_{x+\hat 2,1} - 2 \pi \tilde n _{x+\hat 2,1}}} \nb \\
\ea
This can be compactly written as
\ba
Z &=&  \frac{1}{R} \int D \phi   D  \tilde \phi  \sum_{\pg{\tilde n}} e^{-\frac{1}{4 \pi R^{2} } \sum_{\text{links}} \pr{\D \tilde \phi - v - 2 \pi \tilde n}^{2} - \frac{i}{2 \pi} \sum_{{\rm plaq}}  \pr{ \D \phi-a} \cup \pr{\D \tilde \phi - v - 2 \pi \tilde n} } \times \nb \\
&\times& e^{- S_{\text{sp}}} e^{i \sum_{{\rm plaq}} \pr{\phi \cup \tilde N + N \cup \tilde \phi}}  e^{-i \sum_{{\rm plaq}}  \frac{1}{2 \pi} \D \phi \cup v}  \ .
\ea
As a result, the T-dual action is
\ba
S &=& S_{{\rm bas}} +  \frac{1}{4 \pi R^{2} } \sum_{\text{links}} \pr{\D \widetilde \phi -v - 2 \pi  \widetilde n}^{2} - i \sum_{{\rm plaq}} \pr{  \D \phi \cup  \tilde n  +   \phi \cup \widetilde N} \nb \\
&-& i \sum_{{\rm plaq}} \pq{\frac{1}{2 \pi} A \cup \pr{\D \widetilde \phi - v - 2 \pi \widetilde n}  + N \cup \widetilde \phi}  - \frac{i}{2 \pi} \sum_{\g} \pr{ \D  \phi_{\g} - 2 \pi n_{\g} }  \cup \widetilde \phi   \ . 
\ea
After an integration by parts, we finally obtain 
\ba
S &=& S_{{\rm bas}} +  \frac{1}{4 \pi R^{2} } \sum_{\text{links}} \pr{\D \widetilde \phi -v - 2 \pi  \widetilde n}^{2} + i \sum_{{\rm plaq}}  \phi \cup \pr{ \D  \widetilde n - \widetilde N} \nb \\
&-& i \sum_{{\rm plaq}} \pq{\frac{1}{2 \pi} A \cup \pr{\D \tilde \phi - v - 2 \pi \tilde n}  + N \cup \tilde \phi} \nb \\ 
&-& \frac{i}{2 \pi} \sum_{\g} \pq{ \pr{ \D  \phi_{\g} - 2 \pi n_{\g} }  \cup \widetilde \phi + 2 \pi \phi_{\g} \cup \widetilde n}  \ , \q 
\ea
where we have indicated with the label $\gamma$ that the cochains are fixed at the boundary by the boundary conditions.
We started with 
\ba
\label{eq:NLSMactioncupHcomp}
S &=& S_{{\rm bas}}  + \frac{R^{2}}{4 \pi } \sum_{\text{links}} \pr{\D \phi - 2 \pi n - a}^{2} + i \sum_{{\rm plaq}} \pr{ \D n - N} \cup  \tilde \phi \nb \\
&+& i \sum_{{\rm plaq}} \pq{\frac{1}{2 \pi} \pr{\D  \phi - 2 \pi  n} \cup v  - \phi \cup \widetilde N} \ .
\ea
We thus see that T-duality exchanges
\be
a \leftrightarrow v \ , \q \q \q N \leftrightarrow \widetilde N \ .
\ee

\newpage
%%%%%%%%%%%%%%%%%%%%%%%%%%%%%%%%%%%%%%%%%%%%%%
%%%%%%%%%%%%     Bibliography     %%%%%%%%%%%%
%%%%%%%%%%%%%%%%%%%%%%%%%%%%%%%%%%%%%%%%%%%%%%
\bibliography{biblio}
\bibliographystyle{JHEP}

\end{document}